\documentclass{article}
\usepackage{graphicx} 
\usepackage[affil-it]{authblk}
\usepackage[utf8]{inputenc}
\usepackage[english]{babel}
\usepackage{titlesec}
\usepackage[margin=0.5in]{geometry}
\usepackage{mathtools}
\usepackage{enumerate}
\usepackage{enumitem}
\usepackage{lineno} 
\usepackage{adjustbox}
\usepackage{multirow}
\usepackage{chngcntr}
\usepackage{titlesec}
\usepackage[utf8]{inputenc}
\usepackage{mathrsfs}
\usepackage{lscape}
\usepackage{cite}
\usepackage{cite}

\usepackage{enumitem}
\usepackage{subfigure}
\usepackage{subcaption}

\usepackage[hyphens]{url}
\usepackage{hyperref}
\usepackage{float}
\usepackage[section]{placeins}
\usepackage[dvipsnames]{xcolor}
\usepackage{soul}
\usepackage{amsmath}
\usepackage[labelfont=bf,font=small]{caption}
\usepackage{bm}
\usepackage[title]{appendix}
\usepackage{graphicx}
\usepackage{tabularx}
\usepackage{subcaption}
\usepackage{float}
\usepackage{mathtools}
\usepackage{amsmath,amssymb,amsthm}
\usepackage{oubraces}
\numberwithin{equation}{section}
\usepackage{chngcntr}
\usepackage{mathrsfs}
\usepackage{multirow}
\usepackage{chngcntr}
\usepackage{parskip}
\usepackage{accents}
\usepackage{soul}
\usepackage{orcidlink}

\usepackage[dvipsnames]{xcolor}   

\definecolor{dgn}{rgb}{0.0, 0.5, 0.0}

\newcommand{\allmberry}{\color{Mulberry}{}}

\usepackage{lipsum}
\makeatletter
\g@addto@macro{\endabstract}{\@setabstract}
\newcommand{\authorfootnotes}{\renewcommand\thefootnote{\@fnsymbol\c@footnote}}%
\makeatother

\begin{document}
\title{Phase-Field Modeling of Border Cell Cluster Migration in \emph{Drosophila}}
\author{Naghmeh Akhavan \orcidlink{0000-0002-9474-4486}$^{a}$, Alexander George$^{b}$, Michelle Starz-Gaiano\orcidlink{0000-0002-0855-0715}$^{b}$,  and Bradford E. Peercy \orcidlink{0000-0002-8597-2508}$^{a}\footnote{Corresponding author: Email: {\it bpeercy@umbc.edu}}$\\
{\it {\small $^{a}$  Department of Mathematics and Statistics, University of Maryland Baltimore County, MD, 21250, USA}}

 {\it {\small $^{b}$  Department of Biological Science, University of Maryland Baltimore County, MD, 21250, USA}}

\vspace{-1.5cm}
    }
  \date{}
\maketitle
\vspace{1cm}
\begin{center}
    \textbf{Abstract}
\end{center}
\noindent 
Collective cell migration is a fundamental biological process that drives events such as embryonic development, wound healing, and cancer metastasis. In this study, we develop a biophysically informed phase-field model to investigate the collective migration of the border cell cluster in the \emph{Drosophila melanogaster} egg chamber. 
Our model captures key aspects of the egg chamber architecture, including the oocyte, nurse cells, and surrounding epithelium, and incorporates both mechanical forces and biochemical cues that guide cell migration.We introduce the Tangential Interface Migration (TIM) force which captures contact-mediated propulsion generated along interfaces between the border cell cluster and surrounding nurse cells. 
Our simulations reveal three key features of TIM-driven migration that distinguish it from  previous forms of chemotaxis: (1) the necessity of border cell–nurse cell overlap to initiate movement (\emph{i.e.}, border cells cannot move without a nurse cell substrate), (2) motion is tangential to border cell-nurse cell interfaces, and (3) persistent migration even in regions where the spatial slope of chemoattractant is decreasing. Additionally, we demonstrate that with or without geometry-mediated alterations in chemoattractant distribution such as at intercellular junctions we can vary induced migration pauses, independent of mechanical confinement. We capture an experimentally observed transition to dorsal migration at the oocyte with a sustained medio-lateral chemical cue of small amplitude.  The results show how spatial constraints and interfacial forces shape collective cell movement and highlight the utility of phase-field models in capturing the interplay between tissue geometry, contact forces, and chemical signaling.

\section*{Introduction}
Collective cell migration is a fundamental biological process that underlies diverse phenomena such as embryonic morphogenesis, wound healing, immune surveillance, and cancer metastasis~\cite{ridley2003cell, friedl2009collective,kabla2012collective, montell2012group, peercy2020clustered}. In contrast to single-cell motility \cite{mogilner1996cell, mitchison1996actin, bretscher1996getting, pollard2003cellular, mogilner2009mathematics, mogilner2020experiment}, collective migration involves groups of cells that maintain intercellular junctions and coordinate their movement in response to both internal signaling and external environmental cues. Unraveling the principles that govern such coordinated behaviors is crucial not only for understanding normal tissue development but also for identifying how dysregulated migration contributes to pathological conditions, including tumor invasion and metastatic dissemination.

\begin{figure}
    \centering
    \includegraphics[width=1\linewidth]{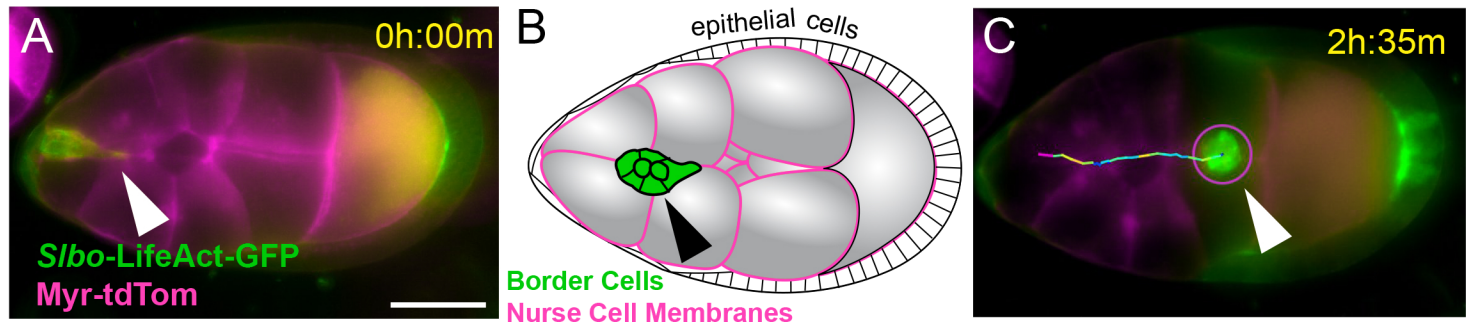}
    \caption{Spatio-Temporal Progression of the Border Cell Cluster. (A, C) Live imaging of border cell cluster migration in the Drosophila egg chamber. Time-lapse images showing the
migration of the border cell cluster (green, Slbo-LifeAct-GFP) through the nurse cell complex toward the oocyte. The cell
membranes are labeled with Myr-tdTomato (magenta). The white arrowhead indicates the position of the border cell cluster at the
time shown at the top right (hours:min).  (B) A cartoon of the egg chamber when border cells (indicated with black arrowhead) have migrated halfway, with cell types labeled.}
    \label{fig:experim}
\end{figure}

The \emph{Drosophila melanogaster} egg chamber is an organized structure that serves as a powerful model for studying collective cell migration. 
Each egg chamber consists of 16 interconnected germline cells, 15 nurse cells and one oocyte, surrounded by a somatic epithelium of follicle cells (Figure~\ref{fig:experim}).
The follicle cells, nurse cells and the oocyte play distinct roles in oogenesis within the egg chamber \cite{king1970ovarian, spradling1993developmental, duhart2017repertoire}.
At the anterior and posterior pole (ends) of the egg chamber are small groups of specialized somatic cells called polar cells, which play a role in signaling and organizing cell migration. A subset of anterior follicle cells is specified as the border cell cluster, which consists of 6-8 migratory epithelial cells that include two non-migratory polar cells and 4-6 motile border cells.
During stages 8 and 9 of oogenesis, the border cell cluster detaches from the anterior epithelium and migrates collectively between the large, immobile nurse cells toward the posterior of the egg chamber, ultimately reaching the oocyte~\cite{king1970ovarian, montell2003border, peercy2020clustered}. 
Border cell migration is guided by tissue topography~\cite{dai2020tissue} and presumed spatial gradients of chemoattractants, including PVF1, {\allmberry a} Platelet Derived Growth Factor and Vascular Endothelial Growth Factor (PDGF/VEGF) family ligand and Epidermal Growth Factor (EGF)-like ligand \cite{duchek2001guidance, mcdonald2006multiple, george2025chemotaxis} secreted from the oocyte \cite{duchek2001guidance}. 
This guidance occurs through receptor-ligand interactions involving receptor tyrosine kinases (RTKs) expressed on the surface of border cells, which activate intracellular signaling pathways that promote polarized protrusion and directional motility \cite{duchek2001guidance2, duchek2001guidance, prasad2007cellular, bianco2007two}.
As demonstrated by George et al.~\cite{george2025chemotaxis}, the effectiveness of this chemotactic response is shaped by tissue geometry, which alters local chemoattractant distribution and modulates migration speed along the path to the oocyte. The physical environment through which the cluster travels is highly constrained, involving complex geometry, deformable neighboring cells, and narrow heterogeneous extracellular spaces between cells that can modulate signal distribution and mechanical resistance.

Mathematical 
models have been instrumental in exploring the mechanisms of collective cell migration. Traditional approaches include agent-based models (ABMs), vertex models, and cellular automata~\cite{jiang2005multiscale, farhadifar2007influence, stonko2015mathematical, camley2017physical, painter2019mathematical}, which offer discrete representations of individual cell behaviors. 
While powerful in capturing rule-based cell interactions, these methods face limitations when addressing continuous and dynamic shape changes, topological rearrangements, and spatially distributed biochemical fields.
In particular, they often require explicit interface tracking or introduce geometric artifacts that hinder the {biophysically appropriate}
simulation of deformation and coordinated group migration within tightly packed tissues.

To address these challenges, we adopt a continuum-based phase-field modeling framework. Phase-field methods represent individual cells as continuous scalar fields, allowing for seamless simulation of complex morphological changes, interface dynamics, and collective rearrangements without explicit tracking of cell boundaries~\cite{provatas2011phase,nonomura2012study, lober2015collisions, najem2016phase, lee2017new, seirin2021extra, wenzel2021multiphase}.
This approach is especially well-suited to multicellular systems with mechanical and chemical couplings, where cells interact through adhesion, volume exclusion, and signaling gradients within a dynamic microenvironment.
{ Although phase-field models have been increasingly applied to biological systems \cite{nonomura2012study, shao2012coupling, palmieri2015multiple, marth2016collective, camley2017physical, moure2021phase, monfared2025multi}, many existing studies simplify key aspects of the tissue architecture.}
For example, the spatial heterogeneity of the extracellular space is often overlooked or homogenized, rather than explicitly represented in the model.
{ Moreover, cell-to-cell variation in size, adhesion properties, and receptor-mediated signaling dynamics is frequently omitted. These simplifications can obscure the complex interplay between tissue geometry, biochemical gradients, and interfacial forces that jointly shape directed collective migration.}

In this study, we develop a biologically informed, multi-cellular phase-field model of border cell cluster migration in the \emph{Drosophila} egg chamber. Our model introduces three key innovations:  (1) a biophysically grounded tangential interface migration (TIM) force that models contact-mediated propulsion along cluster–nurse cell interfaces, enabling migration even in shallow or irregular chemoattractant gradients. 
Each cell type in the system (nurse cells, the oocyte, and the border cell cluster) is represented by a separate phase-field variable. 
The full model is governed by a system of partial differential equations derived from a biophysically motivated energy functional incorporating volume preservation, interfacial tension, and cell-cell adhesion.

\section*{Methods and Results}
{We followed the protocol for culturing \emph{Drosophila melanogaster} stage 9 egg chambers for live imaging of \cite{prasad2007protocol} for the genotypes indicated in the figures}. 
We utilize the multi-cellular phase-field method inspired by \cite{nonomura2012study, seirin2021extra} 
to model collective cell migration in the \emph{Drosophila} egg chamber.

\subsection*{Phase field Model of Egg Chamber Architecture}

\begin{figure}[!htb]
    \centering
    \includegraphics[width=0.75\linewidth]{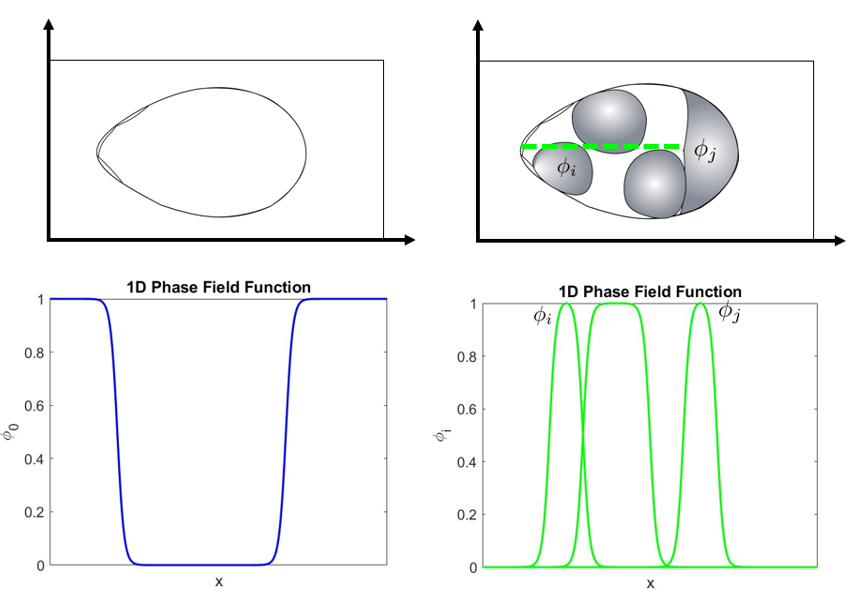}
    \caption{{\bf{Schematic representation of the phase-field model.}} (Left column) The epithelial layer of the egg chamber is considered inside a rectangular region. The egg chamber is described by $\phi_0$ (blue). We consider the inside $\phi_0 = 0$ and outside $\phi_0 = 1$ with the transition region between them as effectively the epithelial layer. (Right column) The gray regions are indicated to nurse cells and oocyte and are considered as phase field $\phi_i$'s. Cross sectional line is taken through the egg chamber (green dashed). 1D phase-field representation of cells along the corresponding cross-section.}
    \label{fig:phis}
\end{figure}

We model the egg chamber, and the cells within, using energy functions.  We include the phase variables for the egg chamber $\phi_0(x,t)$, nurse cells $\phi_m(x,t)$ ($1 \le m \le N=6$), border cell cluster $\phi_c(x,t)$, 
 and oocyte $\phi_{\text{oct}}$, where $x \in \Omega \subset \mathbb R^n$ denotes the area of the system ($n=2$, {two-dimensional} in the top of Figure~\ref{fig:phis}), $t \geq0$ is time. 
 {For simplicity, we represent the entire border cell cluster using a single phase variable, $\phi_c(x,t)$, rather than modeling each individual border cell.}
Each phase variable \(\phi_j\) represents a specific cellular domain (e.g., nurse cells $j=m$, oocyte $j=\mbox{oct}$, or border cell cluster $j=c$), where \(\phi_j = 1\) denotes the interior of the cell and \(\phi_j = 0\) corresponds to the exterior. In contrast, the variable \(\phi_0\) defines the egg chamber, with \(\phi_0 = 0\) inside the egg chamber and \(\phi_0 = 1\) outside (Figure~\ref{fig:phis}).

In our model, individual cells are represented by phase-field variables that are subject to constraints enforcing volume preservation and structural integrity including during border cell migration through the egg chamber. 
The evolution of each phase-field is governed by a Ginzburg–Landau-type equation \cite{landau1963selected, ginzburg2004nobel}, derived as a gradient flow of the system's free energy functional. 

For simplicity of the model, we consider $\phi_m$ for border cell cluster, nurse cells, and oocyte (\emph{i.e.}, number of cells, $N = 8$). We consider the free energy functions of cells as follows:
\begin{align}\label{eq: E0}
    E_0 = \sum_{m=1}^N \int \left[\frac{\epsilon_m^2}{2} |\nabla \phi_m|^2 + g(\phi_m)\right] d{\bf{x}} + \alpha_m \sum_{m=1}^N (V_m(t)- \bar{V}_m(t))^2,
\end{align}
where $g(\phi_m) = \frac{1}{4} \phi_m^2(1-\phi_m)^2$  is a  double-well function in Ginzburg-Landau equation
and $\bar V_m(t)$ and $\alpha_m$ are the target volume, which we take to be fixed in time for each $m$ and the energy intensity of each cell, respectively.
Note that $V_m(t)$ can be defined as below:
\begin{align}\label{eq: vol}
    V_m(t) &= \int_\Omega h(\phi_m)d{\bf x}, \quad  (1 \le m \le N).
\end{align}
In Eq.~\eqref{eq: vol}, we consider the interpolation $h(\phi)$: 
\begin{align}\label{h function}
    h(\phi)=\phi^2(3-2\phi)=\begin{cases}
        0  &\phi=0,\\
        1  &\phi=1, \\
        \mbox{smooth transition}  &0<\phi<1.
    \end{cases}
\end{align}
{Using $h(\phi)$ rather than $\phi$ more accurately approximates the characteristic (indicator) function of the cell domain in phase field models, improving numerical stability and volume conservation, especially near interfaces, because it minimizes interface smearing errors \cite{bhadak2018phase, riva2025rigorous}. }

The parameter $\epsilon_m$ controls the thickness of the diffuse interface between different phases in the phase field model \cite{nonomura2012study}.
Larger values of $\epsilon_m$ result in {wider transition region, meaning the phase field variable changes more gradually between $0$ and $1$. }
Conversely, smaller values sharpen the interface, {leading to more abrupt transition between phases.}
{While the total volume of each phase is theoretically preserved, wider interfaces may introduce slight numerical deviations in volume, especially if not well resolved by the computational grid. }
{Therefore, choosing $\epsilon_m$ involves a tradeoff between interface resolution and numerical accuracy, and computational efficiency \cite{li2023advances}. }

In our model where the sizes of the different cellular components (nurse cells, the border cell cluster, and the oocyte) vary significantly, the parameter $\alpha_m$ plays a critical role in enforcing volume constraints for each phase field variable $\phi_m$. Specially, the term $\alpha_m \left(V_m(t) - \bar V_m(t)\right)^2$ penalizes deviations from the target volume $\bar V_m(t)$, ensuring that the actual volume $V_m(t)$ remains close to the desired value. A larger $\alpha_m$ results in stricter volume conservation by heavily penalizing volume deviations, while smaller values of $\alpha_m$ allow for greater flexibility in volume variation. This term is crucial for ensuring volume preservation, which is a key aspect of the system's dynamics. 

In the next step, we consider the specific conditions of the egg chamber. 
To {model} 
the system, we need to consider specific conditions that reflect the biological environment of the egg chamber.
{Firstly, each nurse cell and the oocyte are influenced within the egg chamber by mechanical interactions with neighboring cells, particularly the compressive forces exerted by surrounding nurse cells
}
Additionally, the nurse cells and the oocyte must occupy {adjacent but spatially distinct} distinct {domains, avoiding interpenetration.}
{The energy form of repulsion} $E_1$ can be considered as following {where any overlap increases the energy in the system:} 
\begin{align}\label{eq: E1}
    E_1 = \beta_0 \sum_{m=1}^N \int h(\phi_0)h(\phi_m) d{\bf{x}} + \beta(m,n) \sum_{m=1}^N \sum_{n \neq m}^N\int h(\phi_n) h(\phi_m),
\end{align}
where $\beta_0$ (epithelial-cell) and $\beta(m,n)$ (cell-cell) are positive constants and represent the intensities of domain territories \cite{nonomura2012study, moure2021phase}.


Also, the egg chamber is entirely filled with nurse cells and the oocyte, all of which are completely enclosed within the epithelial layer. Therefore, we consider $E_2$ as below
\begin{align}\label{eq: E2}
    E_2 = 
    \alpha_0 \left[ \int (1-h(\phi_0))d{\bf x} - \sum_{m=1}^N V_m(t)\right]^2,
\end{align}
where $\alpha_0$ is the energy intensity constant for cells.

Cell-cell adhesion is facilitated by adhesion molecules like cadherins, which are essential for maintaining cell cohesion and for transmitting mechanical signals between cells \cite{gumbiner2005regulation, takeichi1991cadherin}. 
For instance, the adhesion between nurse cells and the oocyte is critical for the structural integrity of the egg chamber and the coordinated migration of the border cells toward the oocyte \cite{montell2003border}.
We consider the adhesion force between nurse cells and each other and nurse cells and oocyte {where opposing gradients from neighboring phases reduce the total energy}
\begin{align}\label{eq: E3}
    E_3 = \gamma_0 \sum_{m=1}^N \int \nabla h(\phi_0) \nabla h(\phi_m)+ \gamma(m,n) \sum_{m=1}^N \sum_{n \neq m}^N\int \nabla h(\phi_m) \nabla h(\phi_n),
\end{align}
where $\gamma_0, \gamma(m,n)>0$ are the intensity of cell-epithelial adhesion and cell-cell adhesion {mediated by molecules like E-cadherins  \cite{friedl2009collective, graner1992simulation, murray2010modelling}, which are critical for maintaining cluster cohesion}.

The total energy of egg chamber is given by
\begin{align}\label{total energy}
    E = E_0 + E_1 + E_2 + E_3. 
\end{align}
We minimize the free energy function to obtain the stable configuration of the cell cluster. Minimizing the free energy of the system ensures that the border cells cluster moves in a coordinated manner toward a stable configuration.

By taking the functional derivative of Eq.~\eqref{total energy}, with respect to the $\phi_m$, we {obtain the variational force driving the system toward energy minimization. 
}
{This derivative is then used to define a gradient flow, resulting in the time evolution equation for $\phi_m$, where phase equilibrium corresponds to a local minimum of the total energy}
\begin{align}
    \frac{\partial \phi_m}{\partial t} &= -\mu \frac{\delta E}{\delta \phi_m}, \quad (1 \le m \le N), \label{wrt phi_m} 
\end{align}
where $\mu >0$ is the mobility of the phase field.  Evaluating the energy Eq.~\eqref{wrt phi_m} becomes
\begin{align}\label{eq: time evol}
    -\frac{1}{\mu} \frac{\partial \phi_m}{\partial t} =& \epsilon_m^2 \nabla^2 \phi_m + \phi_m(1-\phi_m) \left[\phi_m - \frac{1}{2} + 6 \mathcal F(\phi_m, \phi_0)\right], \quad (1 \le m \le N),
\end{align}
in which 
\begin{align}\label{f function}
\begin{split}
    \mathcal F(\phi_m, \phi_0) =& 2  \alpha_m(V_m(t) - \bar V_m(t)) + \beta_0 h(\phi_0) +\beta(m,n) (\xi -h(\phi_m)) \\
    &+2 \alpha_0 \left[\int (1-h(\phi_0)) - \sum_{m=1}^N V_m)\right]+\gamma_0 \nabla^2h(\phi_0)+\gamma(m,n) \nabla^2(\xi -h(\phi_m)).
    \end{split}
\end{align}
where $\xi = \sum_{m=1}^N h(\phi_m)$.

\subsection*{Initializing the Equilibrium State of the Egg Chamber}

\begin{figure}[h]
    \centering
    \subfigure[]{\includegraphics[width=0.45\textwidth]{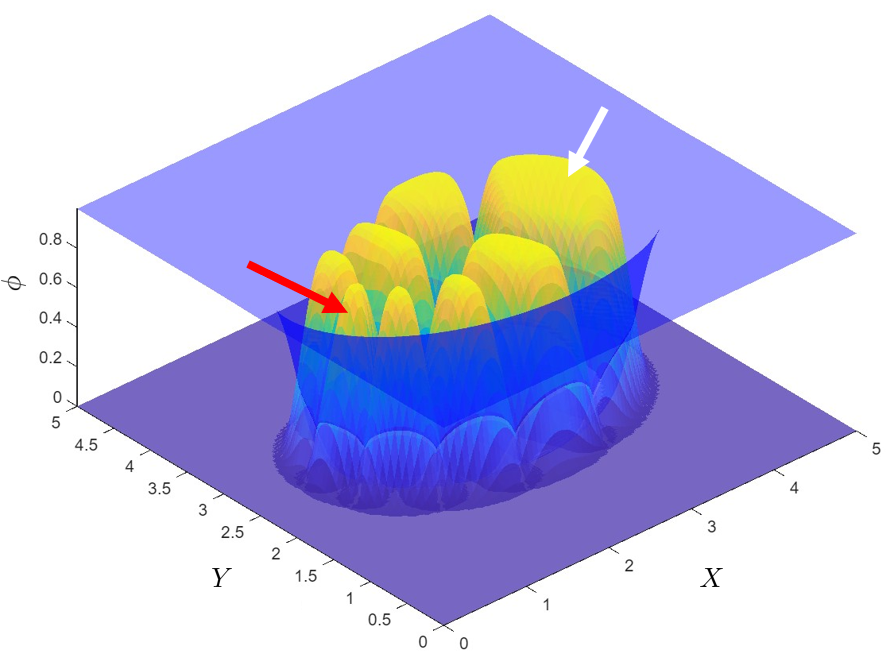}}
    \subfigure[]{\includegraphics[angle=-1,origin=c, width=0.48
    \textwidth]{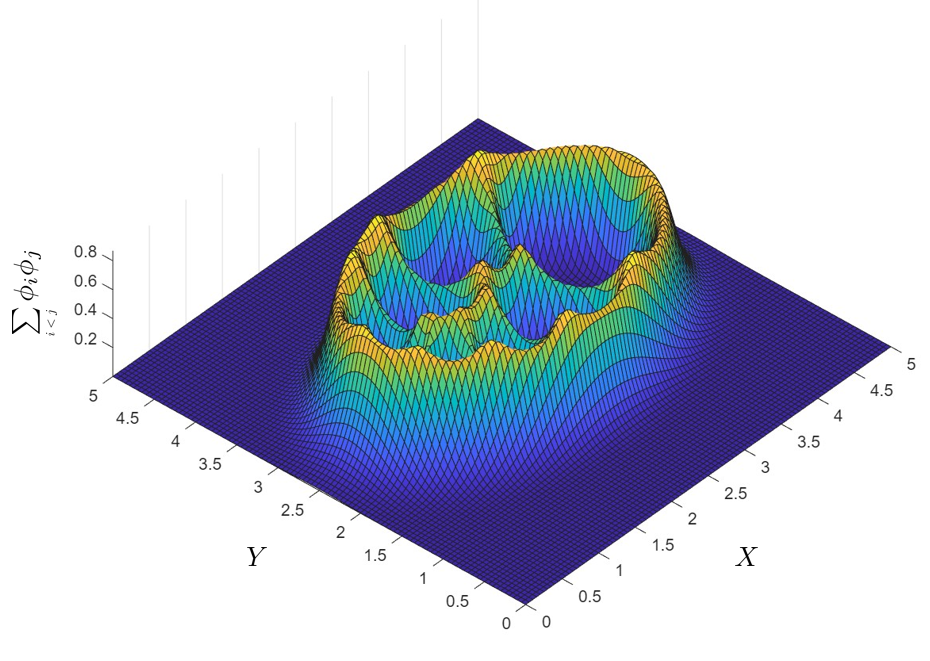}}
    \caption{{\bf{Initial state of the model before border cell cluster migration in the egg chamber.}} The anterior region contains the border cell cluster (red arrows), while the posterior features the oocyte (white arrows), which is larger than other cells. {The model includes six nurse cells that are heterogeneous in size.} (a) {The border cell cluster, nurse cells, and oocyte are shown in yellow. The blue surface represents the epithelial layer phase variable ($\phi_0=0$ inside the epithelium and $\phi_0=1$ outside), defines the boundary of the extracellular space.} {The red arrow shows the position of the border cell cluster at    anterior end, and white arrow indicates the position of the oocyte at posterior end. } 
    (b) Plotting $\sum_{i<j}\phi_i \phi_j$ shows the {interaction are in energy} through which the border cell cluster migrates toward the oocyte. The number of nurse cells is 6. The parameters are set as follow: $\Omega = [0,5] \times [0,5]$, size of spatial grid $h = 0.05$, the time step of $\Delta t = 0.05$, $\epsilon^2 = 0.001, 0.001, 0.0005$ for nurse cells, oocyte, and cluster, respectively. Also, $\alpha_m = 100$ for all $m$, $\beta_0 = 0.9$, $\eta_0 = 0.007$ for epithelial layer. The adhesion intensity  $\beta(1,1)= \beta(1,2)=\beta(1,3)=\beta(2,1)=\beta(3,1) = 0.25$, $\beta(2,3)=\beta(3,2)=0.3$ and $\beta(2,2)=\beta(3,3)=0$. The repulsion intensity $\eta(1,1)= 0.003, \eta(1,2)=\eta(2,1) = 0.004$, $\eta(1,3)=\eta(3,1)=0.008$, $\eta(2,3)=\eta(3,2)=0.005$, and $\eta(2,2)=\eta(3,3)=0$.}
    \label{fig:3d egg try1}
\end{figure}

We define a two-dimensional square computational domain, $\Omega = [0,5]\times [0,5]$, uniformly discretized into an $M \times M$ grid, where the constant mesh size is given by $h = \frac{5}{M+1}$. Time evolution is computed using an explicit time-stepping scheme with uniform step size $\Delta t$. The parameter values $(\alpha_m, \gamma(m,n), \beta(m,n))$ are chosen to reflect adhesion and volume constraints that produce biologically appropriate balance.  These parameters are given in figure captions. Code for simulations is implemented in MATLAB R2022a, with a mesh size of $h=0.05$ and a time step of $\Delta t = 0.05$.
We use the finite difference method to solve Eq.~\eqref{eq: time evol} with Neumann boundary conditions, which are outside of the egg chamber domain. 
 We initialize the system by starting with small circular phases and running the simulation until $t=2000$ and the system reaches an equilibrium where all cells have achieved their corresponding target volumes (Figure~\ref{fig:3d egg try1}).

\section*{Chemoattractant Concentration induces migration}

To investigate the impact of the spatial distribution of chemoattractants within the \emph{Drosophila} egg chamber, we utilize a simple piecewise hyperbolic trigonometric function  that captures secretion, diffusion, and degradation of chemoattractant molecules. These molecules are secreted from the anterior surface of the oocyte and form gradients that guide the directed migration of the border cell cluster through the surrounding nurse cell complex \cite{duchek2001guidance, mcdonald2003pvf1, duchek2001guidance2, mcdonald2006multiple, dai2020tissue}.



\begin{figure}[ht!]

     \begin{center}
        \subfigure[]{%
            \label{fig:first}
            \includegraphics[width=0.45\textwidth]{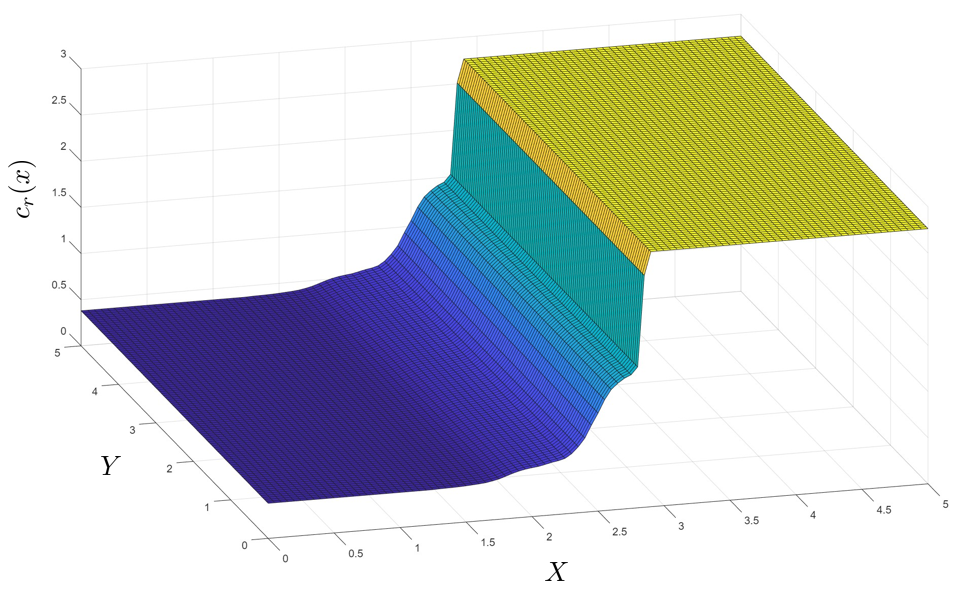}
        }%
        \subfigure[]{%
           \label{fig:second}
           \includegraphics[width=0.45\textwidth]{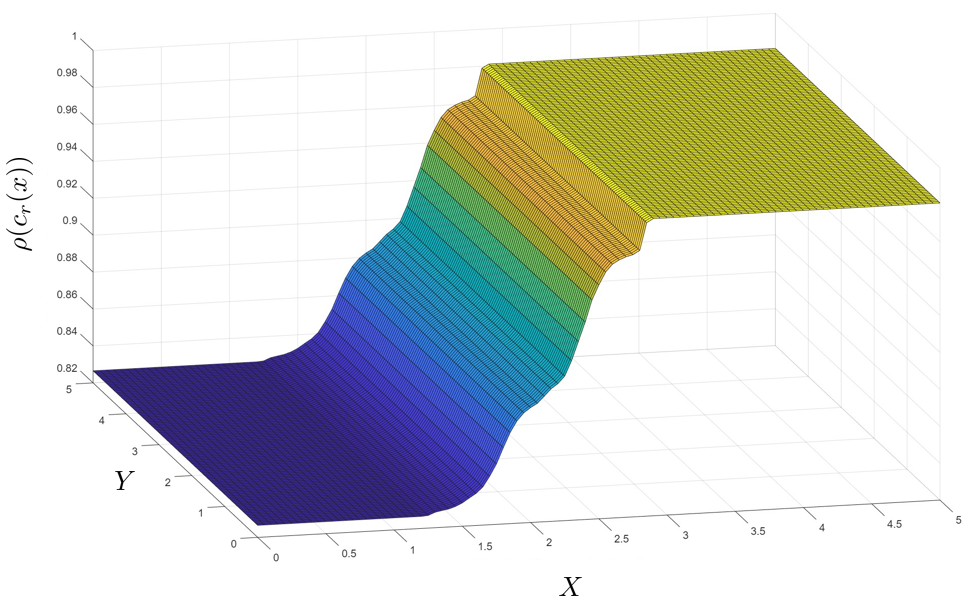}
        }\\ 
        \subfigure[]{%
            \label{fig:third}
            \includegraphics[width=0.3\textwidth]{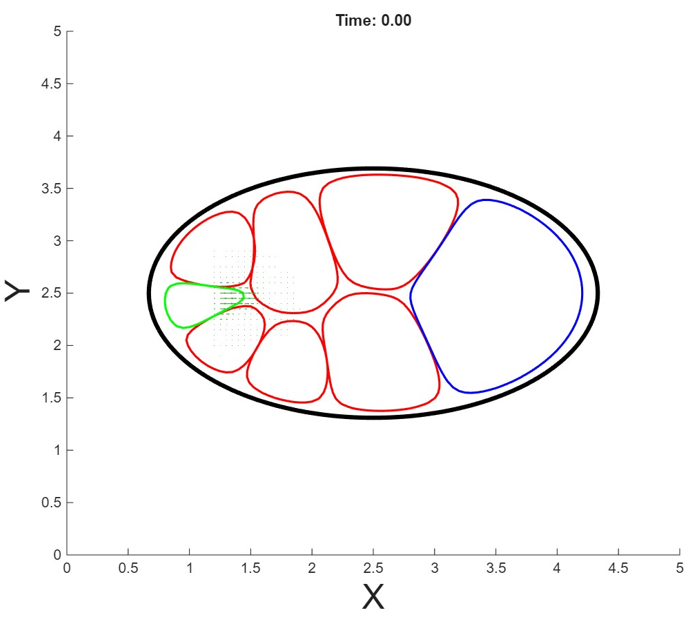}
        }%
        \subfigure[]{%
            \label{fig:fourth}
            \includegraphics[width=0.3\textwidth]{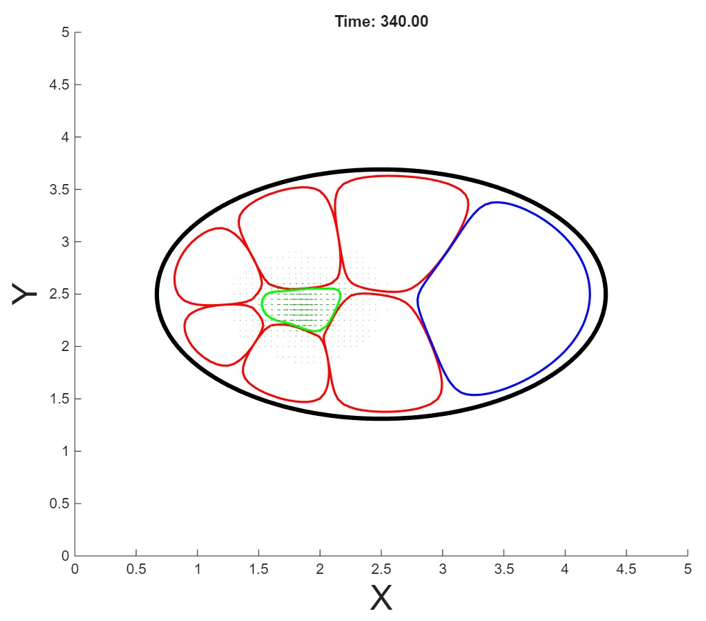}
        }%
        \subfigure[]{%
            \label{fig:fourth}
            \includegraphics[width=0.3\textwidth]{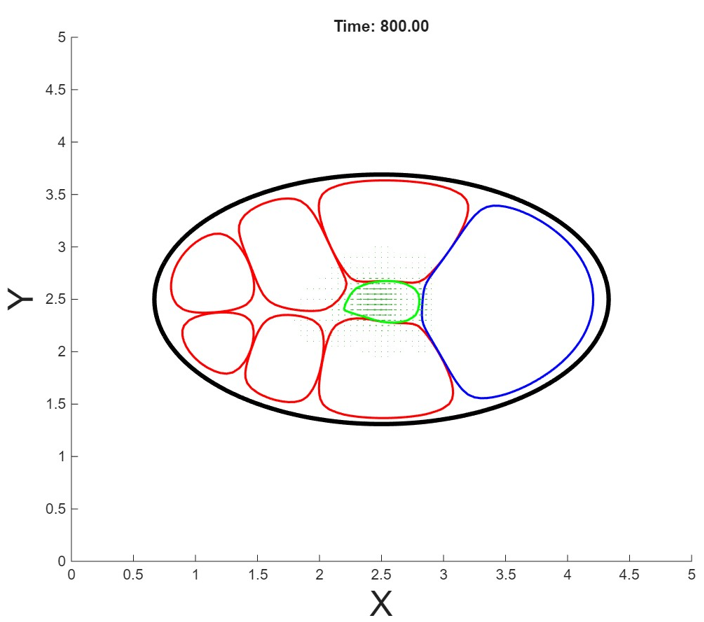}
        }%
    \end{center}
    \caption{%
    {\bf{Impact of traditional chemical force term ($F_{\text{chem}}$).}}
        (a) 2D steady-state distribution of chemoattractant concentration $c(x,y)$ in the egg chamber domain.  (b) Receptor activation in response to the chemoattractant concentration increases along the anterior-posterior axis and saturates at posterior near the oocyte. (c)-(e) Time evolution of border cell cluster migration, simulated using the phase field model in Eq.~\eqref{eq: time evol} with $\mathbf{F}_{\text{chem}}$, Eq. (\ref{eq: Fchem}). The border cell cluster (green) initiates at the anterior end and migrates toward the posterior oocyte in response to a chemoattractant gradient. { The green arrows indicate ($\phi_c \nabla c$) in Eq.~\eqref{eq: Fchem}.} Cell boundaries are represented by phase field contours (red for nurse cells, blue oocyte, and black epithelial layer) at the $\phi=0.5$ level.
        The parameters are set as follow: 
        $\mu_c= 0.045$, $\Gamma= 0.01$, $s=1$, and $\ell = 0.05$. The parameters used are consistent with those in Figure~\ref{fig:3d egg try1}.
     }%
   \label{fig:chem}
\end{figure}

In George et al. \cite{george2025chemotaxis} we show that gaps between nurse cells can generate a heterogeneous distribution of chemoattractant along the anterior-posterior axis.  There we used radial symmetry with varying radii assumption to generate a 1D model capturing the 3D volume variation.  Here we assume the mediolateral space between nurse cells as $2r(x)$ (ultimately representing cross-sectional area used in solving Equation (\ref{eq:chemo1}) in the supplement) and construct a 1D distribution with radial variation, $c_r(x)$ which we trivially extend to $c_r(x,y)$ for a way to capture a representation of gap-generated chemoattractant variation (see Figure \ref{fig:chem}(a)). 
At steady state, the chemoattractant forms a spatial gradient with high concentration near the oocyte surface and lower levels toward the anterior side of the egg chamber \cite{duchek2001guidance, george2025chemotaxis}. The border cell cluster senses this gradient through receptor-mediated signaling and polarizes in response to local concentration differences \cite{duchek2001guidance, duchek2001guidance2, montell2003border, prasad2007cellular, bianco2007two, assaker2010spatial}. Although the chemoattractant profile is temporally constant, spatial asymmetry in its distribution drives directed migration: cells at the leading edge of the cluster experience higher chemoattractant levels than those at the rear, resulting in 
coordinated forward motion \cite{cai2014mechanical, montell2012group}.

Receptor-mediated signaling enables the border cell cluster to interpret extracellular chemoattractant concentrations and initiate directed migration. 
We assume receptor activation depends nonlinearly on the local chemoattractant concentration $c$, capturing cooperative effects and saturation behavior. We use receptor dynamics $\rho(c)$ in functional form of \cite{george2025chemotaxis}:
\begin{equation}\label{eq:rhoa_simplified}
\rho(c) = \frac{s c^3}{(c^2 + \Gamma)(c + \ell)},
\end{equation}
where $s$ is the maximal activation level, {$\Gamma$ and $\ell$ control the sensitivity of the response (Figure~\ref{fig:chem} (b)).} 


\subsection*{Traditional Chemoattractant Force}
Previous work has the chemotactic contribution to a phase-field system given by the term
\begin{align}\label{eq: Fchem}
    \mathbf{F}_{\text{chem}} = -\mu_c \nabla \cdot (\phi_c \nabla c),
\end{align}
where $\mu_c$ is the chemoattractant sensitivity for the phase field variable of border cell cluster $\phi_c$, and $c$ denotes the concentration of the chemoattractant \cite{nonomura2012study}. This force shows how the cluster phase variable $\phi_c$ responds to spatial gradients in the chemoattractant concentration. The variation in the vectorfield (divergence) of the phase field, $\phi_c$, and the gradient of the chemoattractant, $\nabla c$, drives the movement of cluster toward higher concentrations of the chemoattractant. The strength of this response is modulated by $\mu_c$, which determines how strongly the cluster is influenced by the chemoattractant. 

We add the chemical force ($\mathbf{F}_{\text{chem}}$) to the right-hand side of the phase-field evolution Eq.~\eqref{eq: time evol}.  Figure~\ref{fig:chem} (c-e) illustrate the migration of the border cell cluster through the extracellular spaces between nurse cells. First, the chemoattractant reaches a steady-state distribution within the extracellular domain. At this stage, the border cell cluster senses the concentration gradient and initiates directed migration. In addition to the chemotactic force induced by the oocyte-secreted chemoattractant, the cluster relies on adhesion and repulsion forces arising from interactions with the nurse cells and the oocyte. As the cluster migrates, it experiences spatially varying mechanical and chemical cues, which influence its speed. Near the oocyte, receptor saturation occurs due to the elevated chemoattractant concentration, resulting in a reduction in migration speed and eventual attachment of the cluster to the oocyte.

\subsection*{Tangential Interface Migratory (TIM) Force}

Experimental studies, including live imaging and adhesion molecule manipulation, indicate that border cell migration within the \emph{Drosophila} egg chamber involves more than chemotactic guidance or direct propulsion through nurse cells. The cluster frequently displays tangential behaviors—such as spreading, sliding, and crawling—along nurse cell surfaces~\cite{prasad2007cellular, aranjuez2016dynamic}. These behaviors are accompanied by dynamic, actin-rich protrusions and transient cell-cell contacts, pointing to a critical role for mechanical interactions at the cluster–nurse cell interface. 
\begin{figure}[h]
    \centering    
    
    \includegraphics[width=0.9\linewidth]{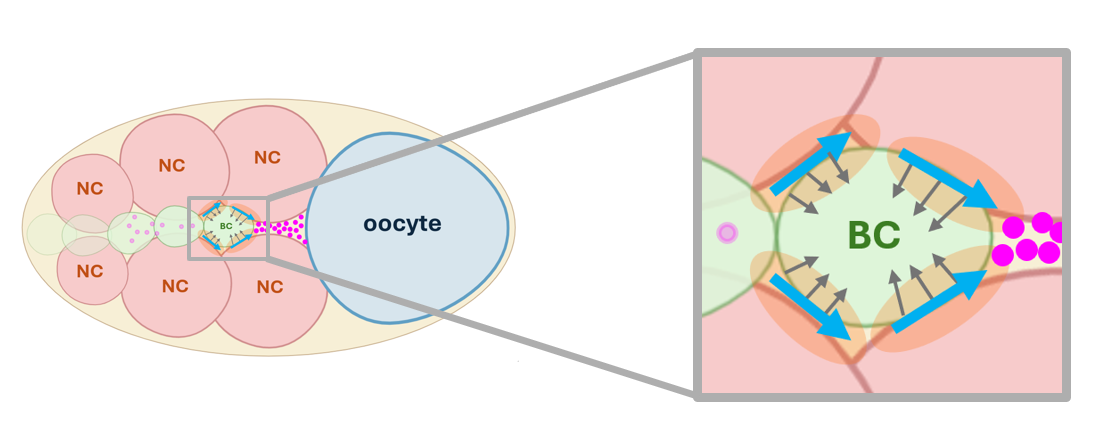}
    \caption{{\bf{Schematic representation of Tangential Interface Migration (TIM) force. }}  The border cell cluster (BC, green) is surrounded by nurse cells (NC, pink), the region of contact between the border cell cluster and nurse cells are shown in orange representing the overlap of phases (border cell cluster phase variable $\phi_c$ and nurse cell phase variables $\phi_j$). The small gray arrows shows the gradient of cluster phase $\nabla \phi_c$ and large blue arrows shows the tangential vectors corresponding the vectors orthogonal to the gradient, ($\nabla \phi_c)^\perp$. 
    Chemoattractant molecules (magenta dots) shows the presumed concentration at front of cluster (posterior side) is more than the back of the cluster (anterior side), which amplifies the movement of cluster from the anterior to the posterior. }
    \label{fig:schematic tim force}
\end{figure}
Prasad and Montell~\cite{prasad2007cellular} observed lateral and rearward protrusions and dynamic positional rearrangements, supporting the idea of contact-mediated motion. Aranjuez et al.~\cite{aranjuez2016dynamic} further showed that border cells generate traction via localized actomyosin contractility and Myo-II enrichment at the interface with nurse cells, highlighting the role of spatially organized mechanical forces in collective migration. These observations collectively support the existence of shear-like, tangential traction forces not captured by normal adhesion or repulsion alone.
To represent this mechanism mathematically, we introduce a novel force term—the \emph{Tangential Interface Migration (TIM)} force—which captures the directionally biased, contact-mediated propulsion generated as border cells use surrounding nurse cells as mechanical substrates for forward migration~\cite{duchek2001guidance, cai2014mechanical}.
\begin{figure}[ht!]
     \begin{center}
        \subfigure[]{%
            \label{fig:first}
            \includegraphics[width=.9\textwidth]{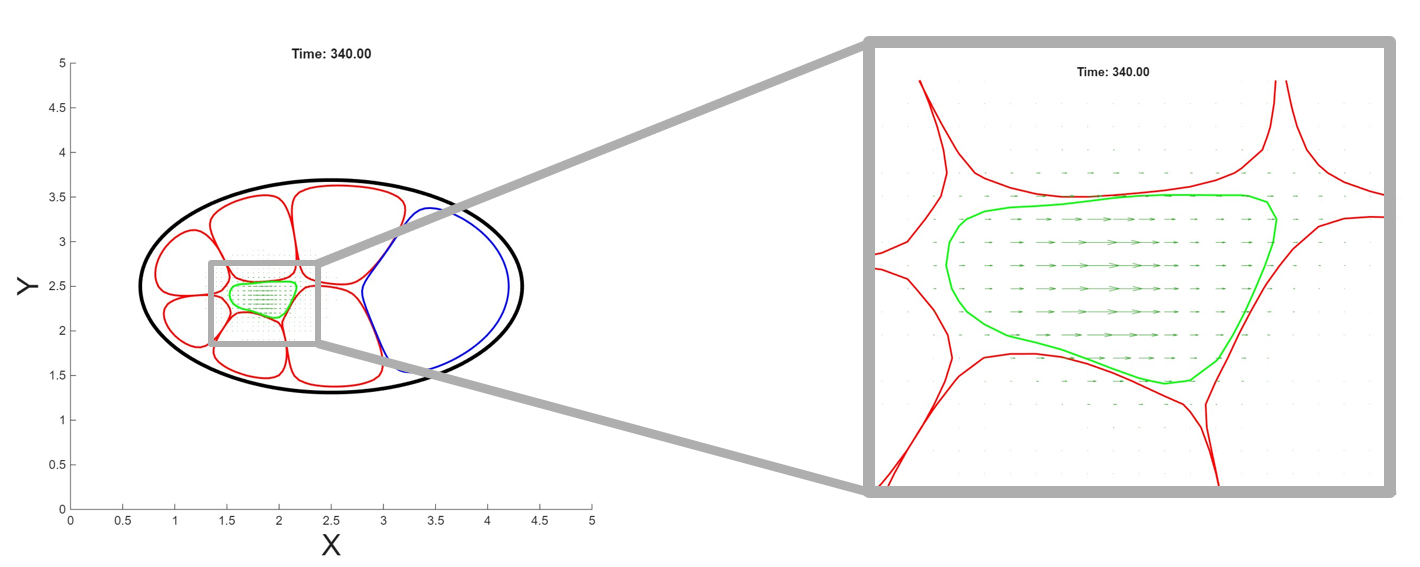}
        }%
        \\ 
        \subfigure[]{%
            \label{fig:third}
            \includegraphics[width=0.9\textwidth]{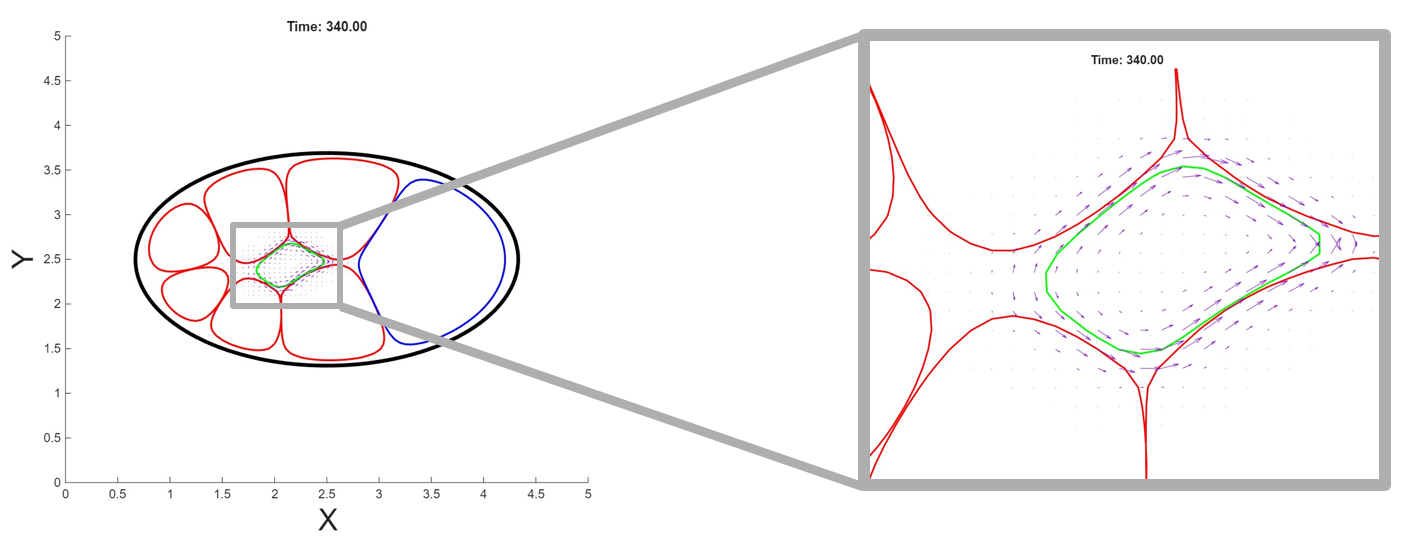}
        }%
%
    \end{center}
    \caption{%
    {\bf{Comparison of $F_{\text{chem}}$ and  $F_{\text{TIM}}$ vector fields.}} Vector fields prior to taking the divergences in each of the chemoatatic forms. (a) Visualization of the chemotactic force vector field ($F_{\text{chem}}$), which directs the cluster toward regions of higher chemoattractant concentration. The magnitude and orientation of the force vectors vary in response to spatial changes in the chemoattractant.
    (b) Visualization of the tangential interface migration force ($F_{\text{TIM}}$),  which is generated by interfacial interactions between the cluster and adjacent nurse cells. The enlarged panel shows tangentially oriented vectors along the cluster boundary, capturing contact-mediated propulsion.
    In both panels migration is shown at the same time $t=340$; nurse cell boundaries are denoted by red; black outlines the epithelial layer of the egg chamber; the blue contour marks the oocyte; and the green region represents the border cell cluster.%
    }
   \label{fig:TIM}
\end{figure}

The TIM force represents contact-mediated, directionally biased traction generated at the cluster-nurse cell interface. Unlike traditional force terms that model repulsion or isotropic adhesion, the TIM force captures the ability of the border cell cluster to ``crawl" or ``climb” along adjacent nurse cells (Figure~\ref{fig:schematic tim force}). We define this force function as below:
\begin{align}\label{eq: tim force}
\mathbf{F}_{\text{TIM}} = -\bar \mu_c \nabla \cdot \left(\rho(c) \, \phi_c \phi_j \, \operatorname{sgn}(\nabla c \cdot \nabla \phi_c^\perp)  \nabla \phi_c ^\perp \right),
\end{align}
where \(\phi_c\) and \(\phi_j\) are the phase field variables representing the cluster and a neighboring nurse cell, respectively. 
The constant parameter $\bar \mu_c$ represents the strength of TIM force. 
The product \(\phi_c \phi_j\) identifies as an overlap, the interfacial region where the border cell cluster is in contact with a neighboring nurse cell. 
The vector \(\nabla \phi_c^\perp\) denotes the tangential direction along the cluster boundary.
The sign function $\operatorname{sign}(\nabla c \cdot \nabla \phi_c^\perp)$ determines the orientation of the tangential force based on the local alignment between the chemoattractant gradient and the cluster boundary. Specifically, it selects the direction along the tangential interface that is aligned with the increase in chemoattractant concentration. This allows the TIM force to promote movement in the direction of higher chemoattractant levels while preserving geometric consistency along the boundary.

{Note that we can also write the TIM force instead of the divergence of a vector field as the inner product of the two vectors fields:
\begin{align}\label{eq: tim force2}
\mathbf{F}_{\text{TIM}} = -\bar \mu_c \nabla \left(\rho(c) \, \phi_c \phi_j \right) \cdot \left( \operatorname{sgn}(\nabla c \nabla \phi_c^\perp)   \nabla \phi_c ^\perp \right),
\end{align}
which shows the gradient alignment of chemoattractactant through the receptor activation function, $\rho$,  with the tangential vector field is key to migration progression.}
{The term \(\nabla \left(\rho(c)\, \phi_c \phi_j\right)\) captures the spatial variation in receptor-ligand interactions along the contact interface, where \(\rho(c)\) represents a receptor-mediated response to the chemoattractant concentration \(c\). Eq.~\eqref{eq: tim force2} captures the alignment between the gradient of receptor-regulated interfacial contact and the local tangential direction of the cluster boundary. The direction of the TIM force is mechanically constrained by cell geometry, while its magnitude is modulated by spatial variations in chemoattractant concentration through receptor-ligand interactions.}

The chemical force {(Eq.~\eqref{eq: Fchem})} on the right-hand side of the phase-field evolution Eq.~\eqref{eq: time evol} is replaced by the TIM force (Eq.~\eqref{eq: tim force}).
Figure~\ref{fig:TIM}(a) shows the spatial localization and orientation of the chemical force during border cell migration. The green arrows in the right panel represent the magnitude and direction of the chemical force, which is spatially localized due to the chemoattractant concentration gradient.
In Figure~\ref{fig:TIM} (b),
the left panel displays the border cell cluster (green interface) in contact with nurse cells (red interfaces), with purple arrows indicating the tangential interface migration vectors localized at the overlapping regions and point from the anterior to the posterior, consistent with the chemoattractant gradient and the cluster’s net migration direction. The right panel magnifies this interaction zone, showing that the TIM force is spatially restricted to areas where the border cell cluster overlaps with nurse cells and aligns tangentially to the cell boundary, reinforcing the notion of contact-guided movement.

\begin{figure}

      \begin{center}
        \subfigure[]{%
            \label{fig:third}
            \includegraphics[width=0.3\textwidth]{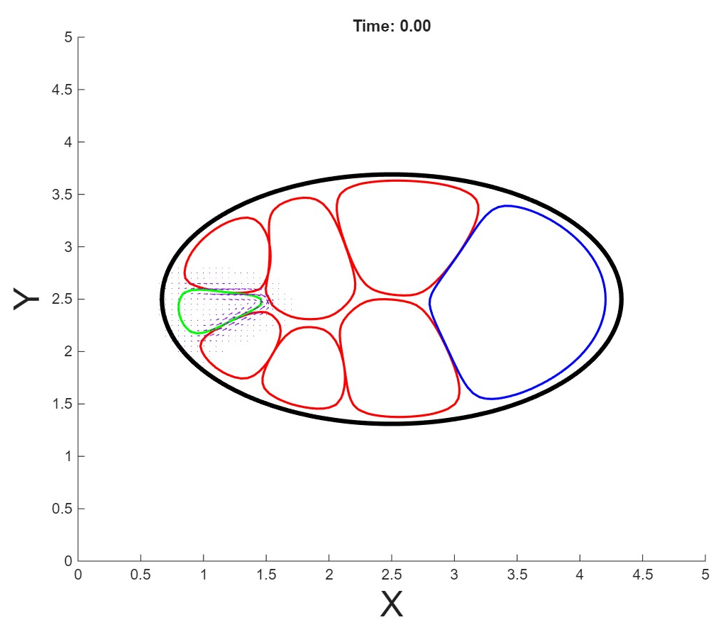}
        }%
        \subfigure[]{%
            \label{fig:fourth}
            \includegraphics[width=0.3\textwidth]{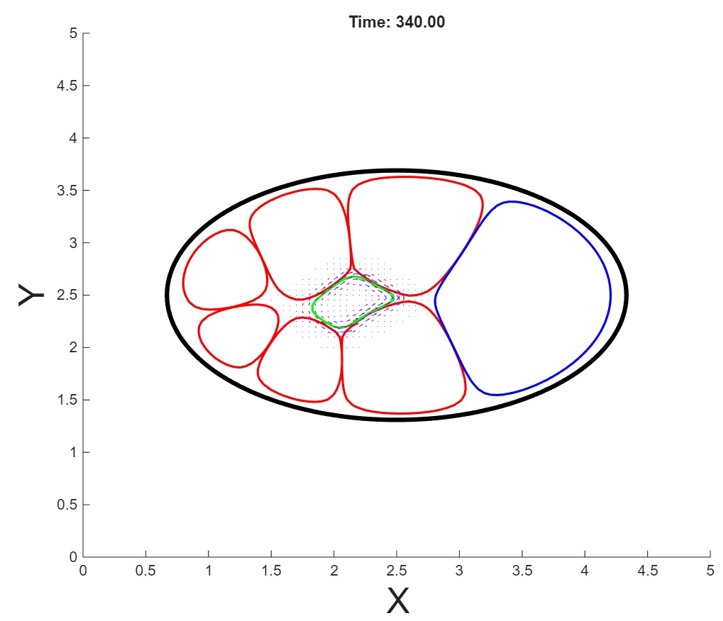}
        }%
        \subfigure[]{%
            \label{fig:fourth}
            \includegraphics[width=0.3\textwidth]{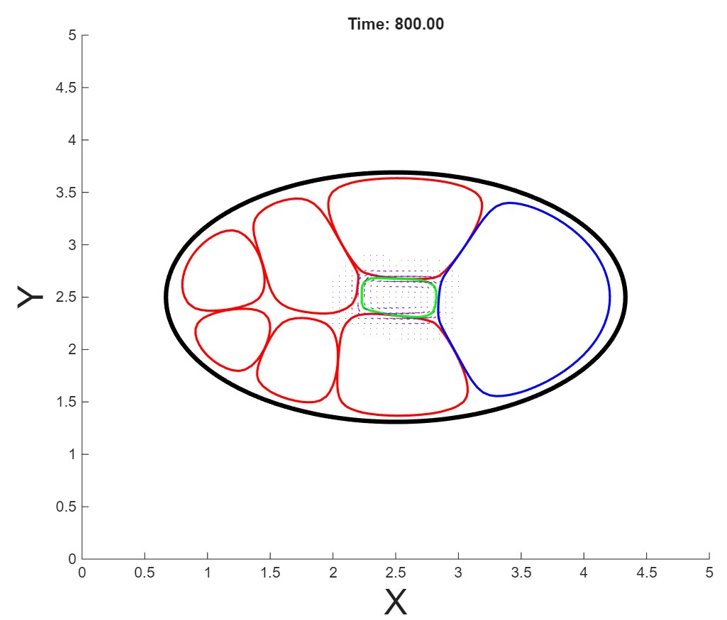}
        }%
    \end{center}
    \caption{%
    {\bf{Impact of Tangential Interface Migration force ($F_{\text{TIM}}$).}} Snapshots of the cell configuration and TIM force vector field at three time points
        (a) $t=0$: The cluster is initially positioned near the anterior of the egg chamber and begins contact-mediated migration along adjacent nurse cell interfaces. Forward movement is initiated as the leading edge of the cluster experiences higher chemoattractant concentration, as indicated by the purple arrow at the front.  (b) $t=340$: The cluster is midway through its migration path, exhibiting tangential traction along adjacent nurse cell boundaries. Vectors indicate directed force aligned with the interfaces.  (c) $t=800$: The cluster approaches the oocyte boundary. At this stage, the front of the cluster senses reduced chemoattractant signaling due to receptor saturation, leading to a gradual decrease in migration speed and eventual attachment to the oocyte. 
     }%
   \label{fig:TIM mig}
\end{figure}

\begin{figure}[ht!]
    \centering
    \subfigure[]{\includegraphics[width=0.46\textwidth]{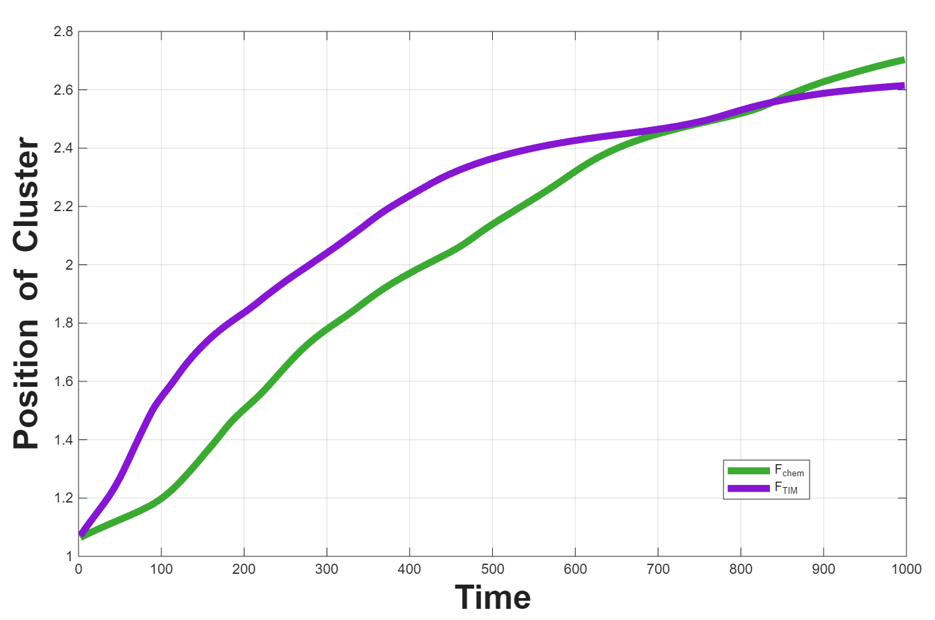}}
    \subfigure[]{\includegraphics[width=0.45
    \textwidth]{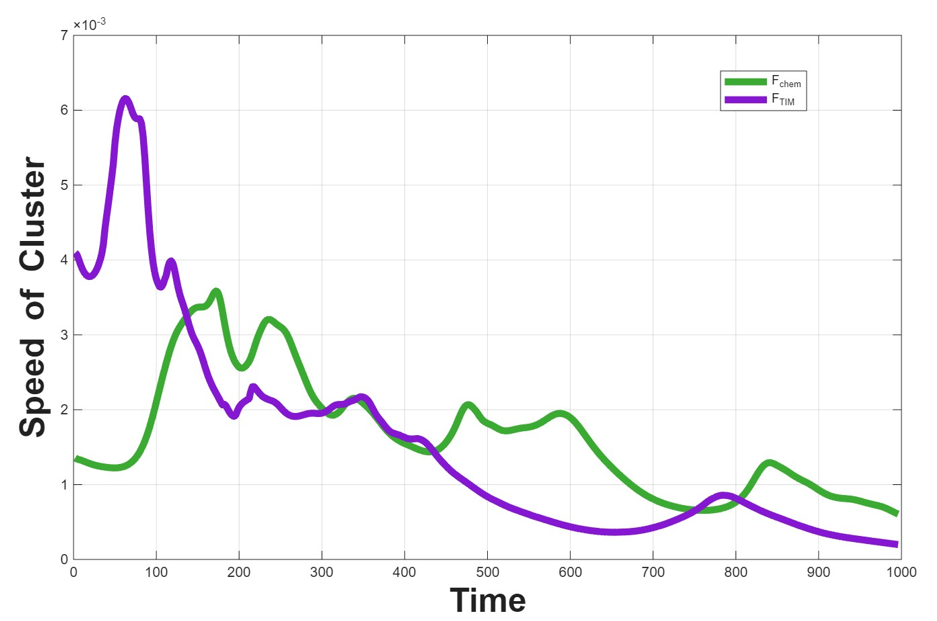}}
    \caption{{\bf{Comparison of border cell cluster migration under $F_{\text{chem}}$ and $F_{\text{TIM}}$.}} (a) Temporal evolution of the center of cluster’s position along the anterior-posterior axis. (b) Speed of the border cell cluster during migration along the anterior–posterior (A–P) axis.}
    \label{fig:pos&speed}
\end{figure}

To further analyze the effect of the TIM force ($\mathbf{F}_{\text{TIM}}$) on migration behavior, we simulate the time evolution of the cluster and track its movement under the influence of TIM. As shown in Figure~\ref{fig:TIM mig} (a-c), the panels present the migration of the cluster progressing through the extracellular space by engaging with and being pushed by adjacent nurse cells. The motion occurs even in the absence of chemoattractant-based force (\emph{i.e.,} $\mathbf{F}_{\text{chem}}=0$), showing the mechanical efficacy of the TIM force. 
{In Figure~\ref{fig:pos&speed} (a), the cluster under TIM force (purple) migrates more rapidly in the early phase and reaches the oocyte boundary earlier compared to the cluster guided by chemotactic force (green), which progresses more gradually. In Figure~\ref{fig:pos&speed} (b) cluster speed over time. The TIM-driven cluster exhibits an early peak in speed, followed by a gradual decline as it approaches the oocyte, likely due to receptor saturation and reduced interfacial traction. In contrast, the chemotactic force leads to more variable speed with multiple transient peaks, reflecting sensitivity to changes in the chemoattractant gradient shaped by the extracellular geometry. }
\section*{TIM‑force reproduces Gurken‑driven dorsal guidance}
{When the cluster is close to the anterior surface of the oocyte it executes a} $\sim$ $10-15 \mu$m ``dorsal turn” toward the germinal vesicle. Genetic studies show that this second phase is driven by a short‑range epidermal growth‑factor–receptor (EGFR) cue that originates from the oocyte {nucleus~\cite{duchek2001guidance, mcdonald2006multiple}.} Gurken (Grk), a TGF‑$\alpha$-like ligand, accumulates as a crescent on the dorsal‑anterior oocyte membrane and activates EGFR on the leading edge of the border‑cell cluster. 
Eliminating EGFR activity in border cells or mutating \emph{grk} abolishes the dorsal movement, but mutating { other guidance factors} does not~\cite{duchek2001guidance2, mcdonald2006multiple}. 

Because Grk protein diffuses only a short distance, its effect is confined to border cells in close proximity to the oocyte surface.
Earlier anterior  to posterior migration, by contrast, relies chiefly on the long‑range PDGF/VEGF ligand PVF1 signaling {through PVR,  and to some extent on other EGFR ligands besides Grk, but these latter proteins are believed to be expressed in spatiotemporal patterns inconsistent with dorsal cues~\cite{mcdonald2006multiple}.}

To produce this two-step chemotactic landscape we prescribe a steady concentration field $c(x,y) = \cosh(1.5 x) + 10^{-4} \cosh(y)$, with $x$ and $y$ denoting the anterior-posterior (AP) and medio-lateral axes, respectively. The AP component $\cosh(1.5x)$ yields a steep PVF1-like  gradient that pulls the cluster through the nurse cell matrix. The dorsal component $10^{-4}\cosh(y)$ introduces a shallow dorsal bias, though present throughout the chamber, that becomes appreciable only after the cluster reaches the oocyte, mimicking the short-range Grk signal. 
\begin{figure}[ht!]
    \centering
    \includegraphics[width=1\linewidth]{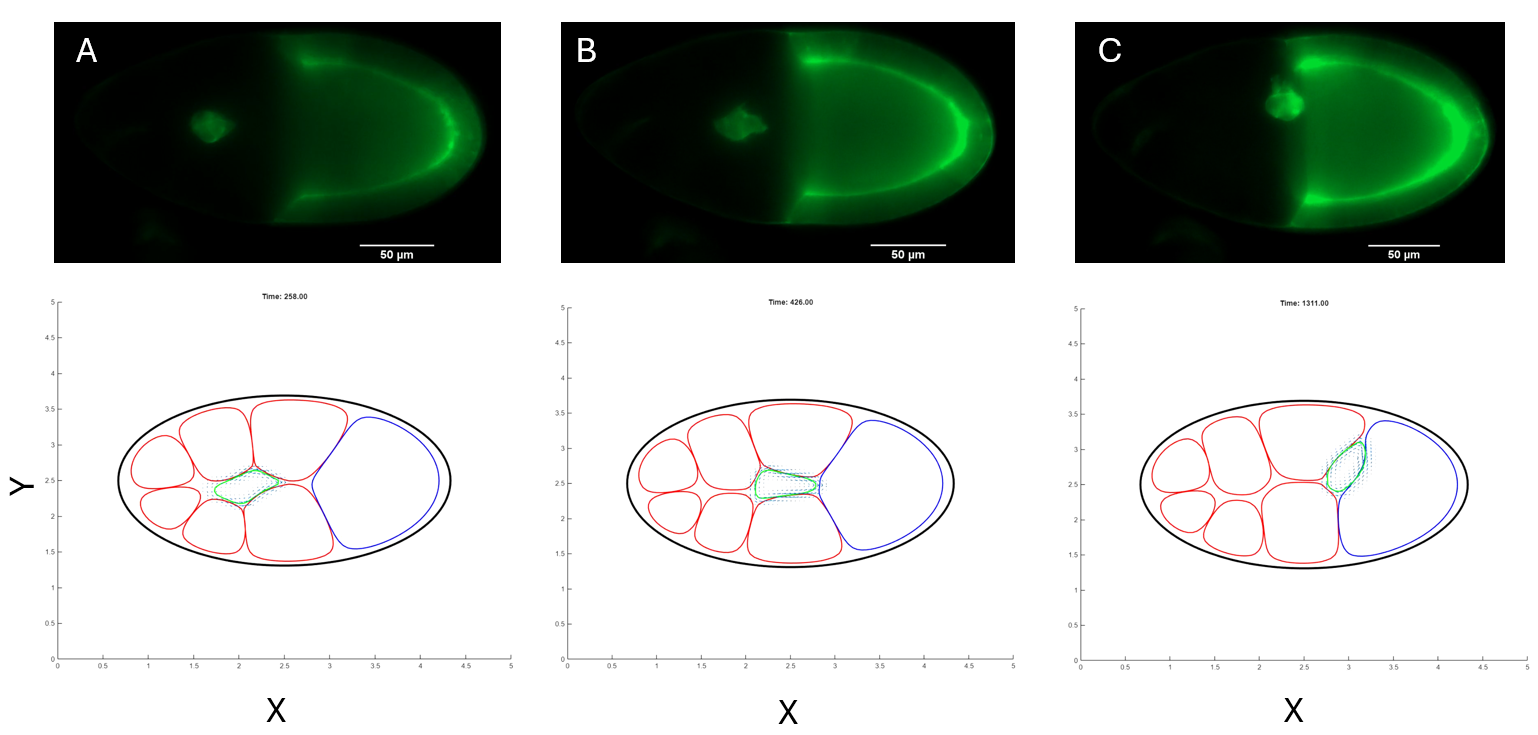}
    \caption{{\bf{Short‑range Gurken cue drives the dorsal turn of the border‑cell cluster: comparison of live imaging and phase‑field simulation.}} Top row: Time-lapse images of the egg chamber expressing a membrane GFP marker (green). {The border‑cell cluster (bright spot) migrates toward the oocyte (A), and as it nears the anterior surface,} reorients dorsally toward the germinal vesicle (B) and continues {towards the dorsal anterior side of the oocyte (C).}
    Bottom row: Corresponding snapshots from the phase‑field model at the indicated simulation times. The arrows depict the TIM force, which is proportional to the local chemoattractant gradient $c(x,y)$. The shallow dorsal component of $c$ becomes appreciable once the cluster contacts the oocyte, redirecting TIM stresses upward and reproducing the experimentally observed dorsal excursion without additional parameter tuning.}
    \label{fig:dorsal migration}
\end{figure}
Within our framework the TIM force converts local chemoattractant gradients into interfacial traction. Consequently, as soon as the cluster contacts the oocyte, the now-dominant dorsal gradient redirects the tangential stresses dorsally, recreating the experimentally observed dorsal excursion Figure~\ref{fig:dorsal migration}. No additional parameter tuning is required, supporting the view that a localized Gurken cue, together with interfacial mechanics, is sufficient to drive the dorsal turn.
\section*{Discussion}
The coordinated migration of the border cell cluster in the \emph{Drosophila} egg chamber offers a powerful model for dissecting the biophysical principles underlying collective cell motility. 
While chemoattractant signaling is known to play a dominant role in guiding migration, recent experimental observations suggest that mechanical interactions with the surrounding tissue also influence both the directionality and dynamics of the cluster’s movement. 
In this study, we developed a multi-cellular phase-field model that captures key aspects of this process, including extracellular chemoattractant dynamics, receptor-mediated signaling, and contact-mediated mechanical forces.

One of the central contributions of our work is the introduction of the Tangential Interface Migration (TIM) force, which models the ability of the border cell cluster to generate propulsion by engaging tangentially with the surfaces of adjacent nurse cells. 
This force is motivated by live-imaging studies showing lateral protrusions, cell-on-cell sliding, and traction generation at the cluster–nurse cell interface \cite{prasad2007cellular, aranjuez2016dynamic}. 
Our simulations demonstrate that TIM force alone 
is sufficient to drive forward migration through confined spaces. This supports a view in which chemical and mechanical cues act in parallel or even redundantly to ensure robust guidance under variable tissue conditions.

The chemoattractant distribution in our model is shaped by the geometry of the extracellular space, which acts as the domain through which the signal diffuses and degrades. Variations in the width and structure of this space—particularly at cell-cell junctions and intersections—can significantly alter the shape and steepness of the resulting gradient. As shown in George et al.~\cite{george2025chemotaxis}, local expansions or intersections in the extracellular domain lead to a reduction in gradient steepness due to increased spatial dilution and slower accumulation of the signal. This effect diminishes the directional cue available to migrating cells, resulting in a measurable decrease in migration speed within these regions. Our model reinforces the idea that not only the source and decay rates, but also the geometry of the extracellular environment, play a critical role in shaping effective chemotactic guidance.


Importantly, our results suggest that the border cell cluster’s ability to interpret shallow or spatially irregular gradients depends on 
its capacity to generate traction at interfaces. When the TIM force is active, the cluster exhibits enhanced directional persistence and cohesive movement { compared to simulations driven by chemoattractant gradient alone.}
This aligns with prior work showing that actomyosin contractility and E-cadherin-mediated adhesion at the cluster periphery are necessary for coordinated migration \cite{cai2014diverse}. The TIM force provides a {biophysically based} 
framework to account for these contact-based behaviors in a continuum model, offering an alternative to purely chemotactic guidance under certain conditions.


Our results demonstrate that the Tangential Interface Migration (TIM) force offers several advantages over {classical chemotactic force such as $F_{\text{chem}}$.}

{First, TIM requires direct contact between the border cell cluster and the nurse cells to initiate movement, migration does not occur without sufficient overlap at the interface, reflecting the physical necessity of a substrate for generating tangential traction.
Second, TIM drives motion tangential to the border cell–nurse cell interface, capturing mechanical behaviors such as sliding and wrapping, as observed in live imaging studies but not modeled by classical chemotactic approaches.
Third, TIM supports persistent, cohesive migration even in regions where the chemoattractant gradient is weak or decreasing, conditions under which the chemotactic force $F_{\text{chem}}$ can lead to stalled or erratic movement.} {For example, if the chemoattractant distribution has negative concavity, $F_{\text{chem}}$ may stall or even go negative though the chemoattractant gradient may still be positive.  This cannot happen with $F_{\text{TIM}}$.}
Together, these characteristics make TIM a robust and biophysically grounded mechanism for guiding collective migration through the densely packed and heterogeneous environment of the nurse cell complex.

Overall, this work advances our understanding of how border cell clusters integrate chemical and mechanical information during migration. The TIM force offers a biologically grounded mechanism for contact-mediated movement that complements existing models of chemotaxis. More broadly, our results illustrate the value of phase-field methods for modeling migration in structured tissues, where cell-cell interfaces and spatial constraints are critical determinants of migratory behavior.

\section*{Acknowledgments}
We acknowledge funding from NSF DMS \#1953423 to B.E.P. and M.S.-G., NSF IOS \#2303587 to M.S.-G.

\section*{Supplementary Material}

Supporting material can be found online at \url{https://github.com/Naghmeh-Akhavan/Phase-Field-Modeling}.

\bibliographystyle{ieeetr}
\bibliography{sample.bib}


\newpage
\section*{Supplementary}\label{sec: supp}

\subsection*{Impact of Extracellular Geometry on Chemoattractant Distribution and Force Dynamics}

To understand how extracellular tissue architecture affects border cell migration, we investigate the impact of domain geometry on steady-state chemoattractant concentration and the resulting migration forces. In particular, we compare two modeling approaches: one that assumes uniform extracellular geometry (constant cross-sectional area), and another that incorporates spatial variation in extracellular space by modeling a position-dependent cross-sectional area along the anterior-posterior (A–P) axis of the egg chamber.

Our goal is to evaluate how these two chemoattractant profiles influence the behavior of the chemotactic force ($F_{\text{chem}}$) and the tangential interface migration force ($F_{\text{TIM}}$). Because $F_{\text{chem}}$ depends directly on spatial gradients in chemoattractant concentration, it is sensitive to local variations introduced by domain geometry. 
In contrast, $F_{\text{TIM}}$ arises from interfacial mechanical interactions and is indirectly affected through changes in cluster position and geometry rather than the gradient itself. By analyzing migration patterns under each concentration model, we demonstrate that geometrically realistic extracellular domains can significantly weaken chemotactic cues, leading to slower or more erratic movement under $F_{\text{chem}}$, while $F_{\text{TIM}}$ remains effective in maintaining directed migration.

By assuming a positive cross-sectional area $A(x)>0$, we obtain the strong form of the equation:
\begin{align}\label{eq:chemo1}
 \begin{split}
    &\frac{\partial c}{\partial t} = \frac{1}{A} \frac{\partial}{\partial x} \left(DA(x) \frac{\partial c}{\partial x}\right) - k c, \\
    &-D A(x)\frac{\partial c}{\partial x}|_{x=L} = -\sigma, \\
    &-D A(x)\frac{\partial c}{\partial x}|_{x=0} = 0,
 \end{split}
\end{align}
where $L$ is the length of the egg chamber, and the boundary conditions reflect no flux at $x=0$ and secretion at $x=L$.
The steady state solution of Eq.~\eqref{eq:chemo1} is shown in Figure~\ref{fig:chem} (a).

By considering $A(x) \equiv 1$, in Eq.~\eqref{eq:chemo1}, the steady-state solution of this equation with boundary conditions  is given by
\begin{align}\label{concentration solution}
c^*(x) =\frac{\sigma}{D \sqrt{\frac{k}{D}} \sinh(\sqrt{\frac{k}{D}}L)} \cosh \left(\sqrt{\frac{k}{D}}x\right). 
\end{align}

We extend the steady-state solution $c^*(x)$ uniformly along the lateral $y$-direction. Specifically, we define the 2D concentration field as $c(x,y) = c^*(x)$, assuming no variation in the $y$-direction, Figure~\ref{fig:chem-cosh}(a).

We apply the same functional form for the receptor-mediated response $\rho(c)$ as defined in Eq.~\eqref{eq:rhoa_simplified} in the main text. 
Figure~\ref{fig:chem-cosh}(b) shows the receptor-mediated response function $\rho(c(x,y))$, which exhibits a slower rate of increase at both the anterior and posterior ends of the domain. In particular, the plateau near the posterior end (larger $x$) reflects saturation of receptor binding as the chemoattractant concentration approaches its maximum near the oocyte.

\renewcommand\thefigure{S\arabic{figure}}\setcounter{figure}{0}  
\begin{figure}[ht!]

     \begin{center}
        \subfigure[]{%
            \label{fig:first}
            \includegraphics[width=0.45\textwidth]{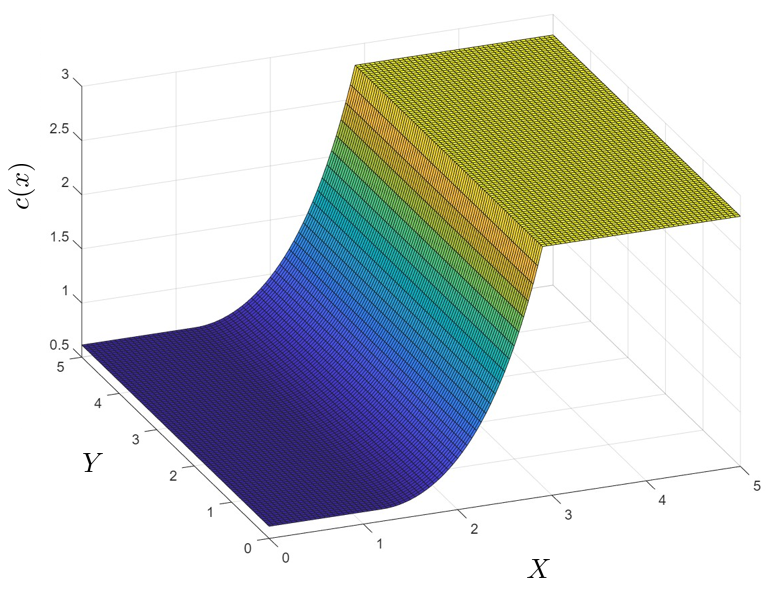}
        }%
        \subfigure[]{%
           \label{fig:second}
           \includegraphics[width=0.45\textwidth]{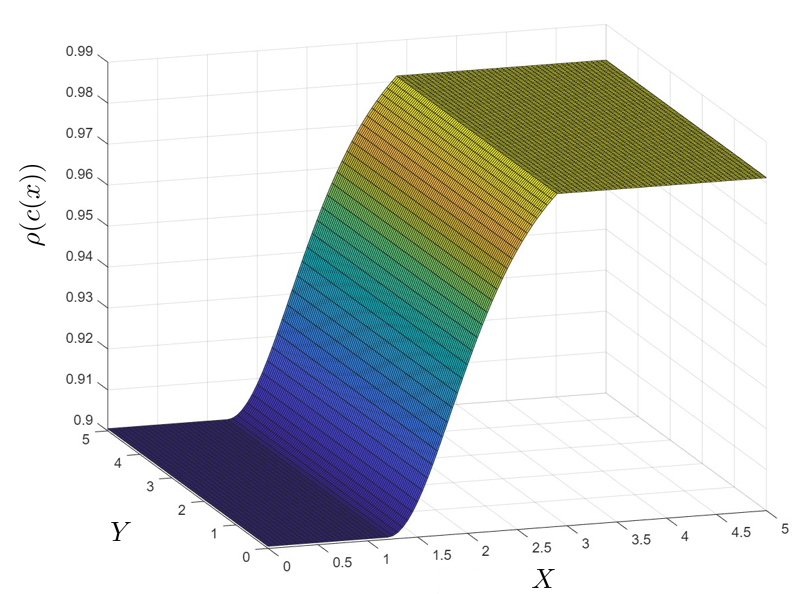}
        }
    \end{center}
    \caption{%
    {\bf{Steady-state chemoattractant concentration $c(x)$ and receptor response $\rho(c)$ in 2D.}} (a) The extended 2D chemoattractant concentration field $c(x,y)=c^*(x)$, where the steady-state solution $c^*(x)$ is derived from $A(x) \equiv 1$. (b) The response curve increases with $x$ but shows slower growth near the anterior and posterior ends, particularly near the oocyte, where receptor saturation occurs due to high ligand concentration.
     }%
   \label{fig:chem-cosh}
\end{figure}


\begin{figure}[ht!]

      \begin{center}
        \subfigure[]{%
            \label{fig:third}
            \includegraphics[width=0.3\textwidth]{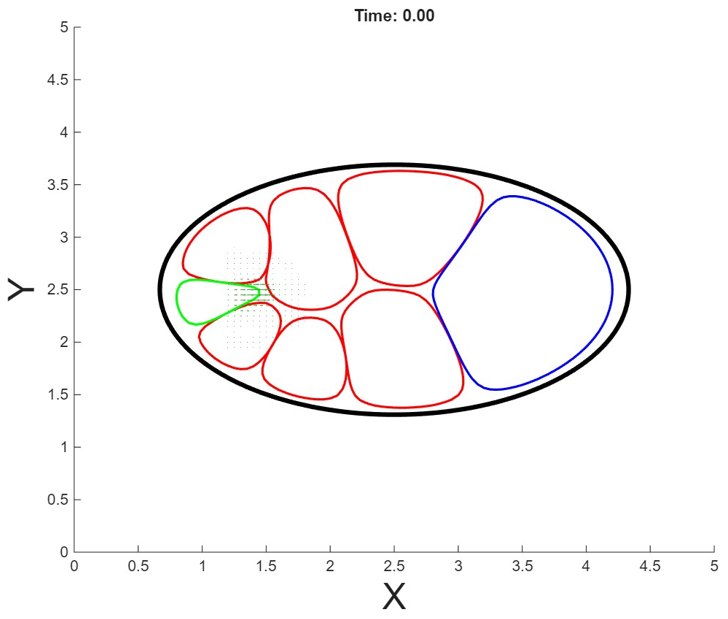}
        }%
        \subfigure[]{%
            \label{fig:fourth}
            \includegraphics[width=0.3\textwidth]{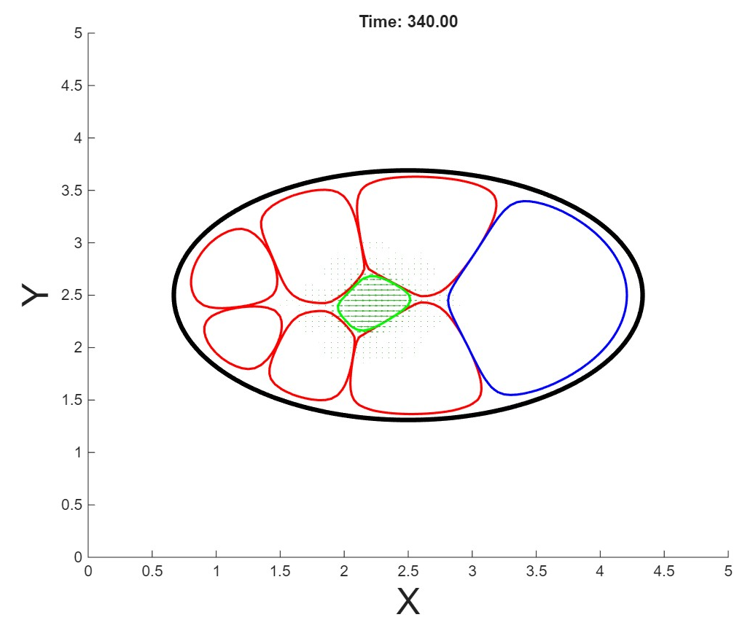}
        }%
        \subfigure[]{%
            \label{fig:fourth}
            \includegraphics[width=0.3\textwidth]{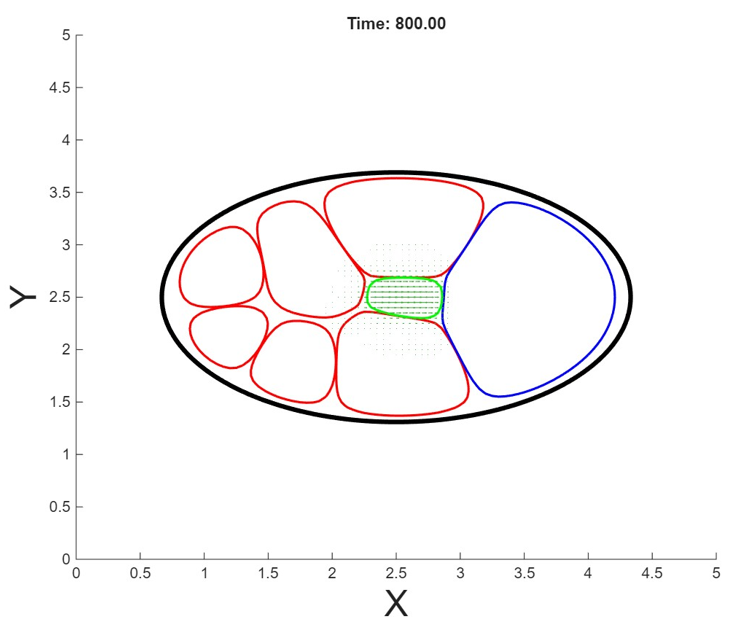}
        }%
    \end{center}
    \caption{%
    {\bf{Migration driven by $F_{chem}$ chemotactic force using uniform chemoattractant concentration $c(x)$.}} Snapshots at three time points show the border cell cluster (green) migrating toward the oocyte (blue) under the influence of $F_{\text{chem}}$. The green vector field represents the gradient of the chemoattractant concentration $\nabla c$, which increases in magnitude along the anterior-posterior axis, guiding forward movement.
     }%
   \label{fig:chem cosh mig}
\end{figure}
We apply the extended chemoattractant concentration field $c(x,y) = c^*(x)$ in the chemotactic force term $F_{\text{chem}}$, which is added to the right-hand side of the phase field evolution equation (Eq.~\eqref{eq: time evol} in main text) to simulate directed migration of the border cell cluster. 

In Figure~\ref{fig:chem cosh mig} (a)-(c), the green vector field in each panel illustrates the direction and magnitude of the chemoattractant gradient $\nabla c$, which determines the direction of $F_{\text{chem}}$. 
Due to the monotonic increase of concentration along the anterior-posterior axis, the gradient vectors consistently point toward the oocyte and gradually increase in magnitude along the migration path. 

\begin{figure}[ht!]

     \begin{center}
        \subfigure[]{%
            \label{fig:first}
            \includegraphics[width=0.47\textwidth]{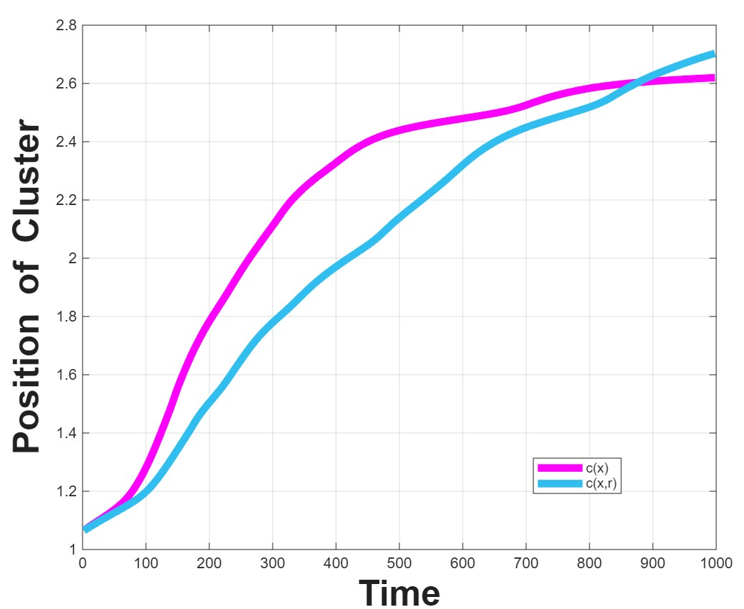}
        }%
        \subfigure[]{%
           \label{fig:second}
           \includegraphics[width=0.45\textwidth]{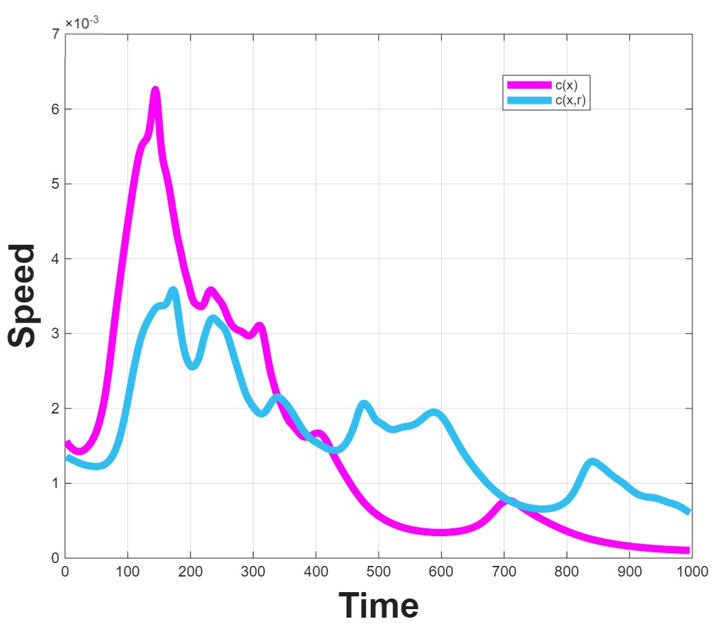}
        }
    \end{center}
    \caption{%
   {\bf{Effect of extracellular space on $F_{chem}$ chemotactic force–driven migration.}} (a) Position of the border cell cluster over time under two chemoattractant profiles: $c(x)$ (magenta), corresponding to a uniform extracellular domain with constant cross-sectional area $A \equiv 1$, and $c_r(x)$ (cyan),  which accounts for spatial variation in extracellular geometry via a non-uniform area $A(x)$. (b)  Corresponding cluster speed profiles. The non-uniform geometry $c_r(x)$ leads to reduced chemotactic gradients in certain regions, resulting in slower migration and more variable speed compared to the uniform case (cyan). 
     }%
   \label{fig:chem cosh and c_r}
\end{figure}

We compare the behavior of border cell migration under two different chemoattractant concentration models: $c(x)$, which assumes a uniform extracellular domain, and $c_r(x)$, which incorporates the influence of spatially varying extracellular geometry through a position-dependent cross-sectional area. The Figure~\ref{fig:chem cosh and c_r} (a)-(b) shows how these differences affect both the position and speed of the migrating cluster.

To investigate how the Tangential Interface Migration (TIM) force responds to chemoattractant profiles independent of spatial heterogeneity, we simulate cluster migration using a uniform chemoattractant concentration based on $c(x)$ from Eq.~\eqref{concentration solution} (Figure~\ref{fig:chem-cosh}), assuming constant cross sectional area. 

\begin{figure}[ht!]

      \begin{center}
        \subfigure[]{%
            \label{fig:third}
            \includegraphics[width=0.3\textwidth]{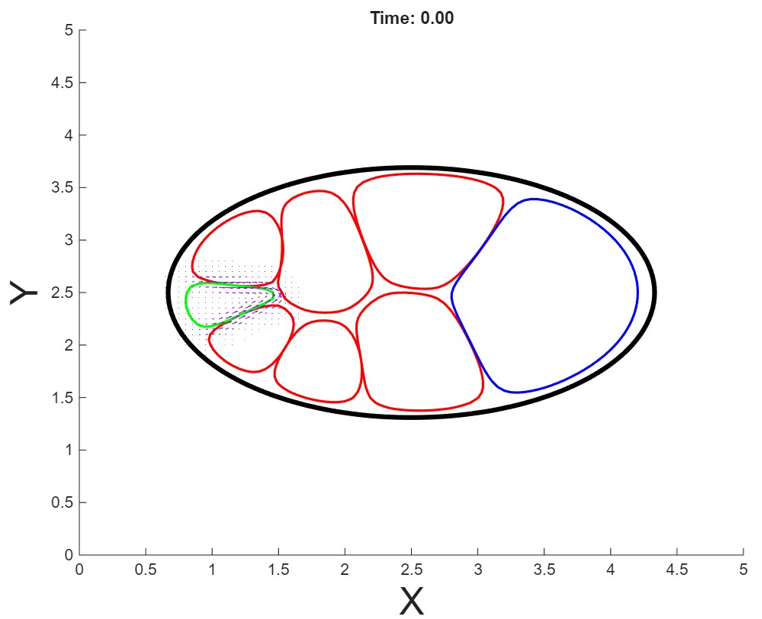}
        }%
        \subfigure[]{%
            \label{fig:fourth}
            \includegraphics[width=0.3\textwidth]{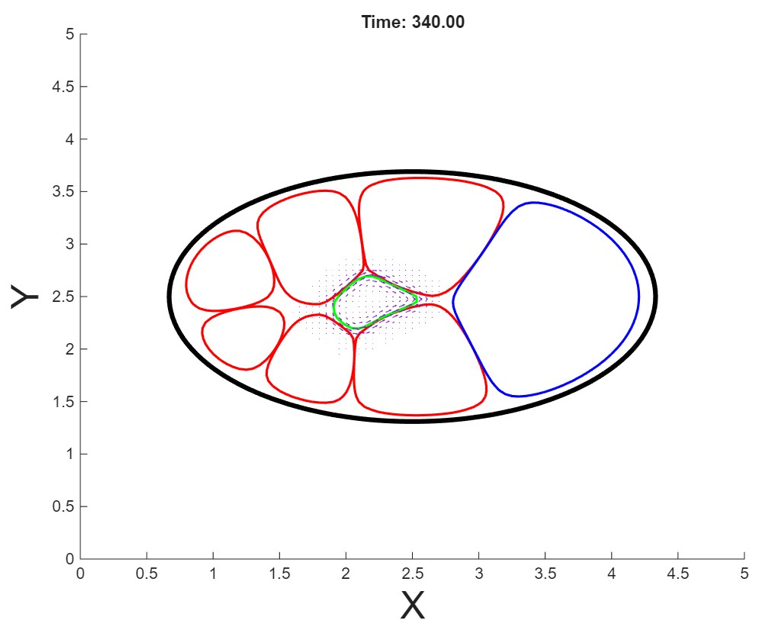}
        }%
        \subfigure[]{%
            \label{fig:fourth}
            \includegraphics[width=0.3\textwidth]{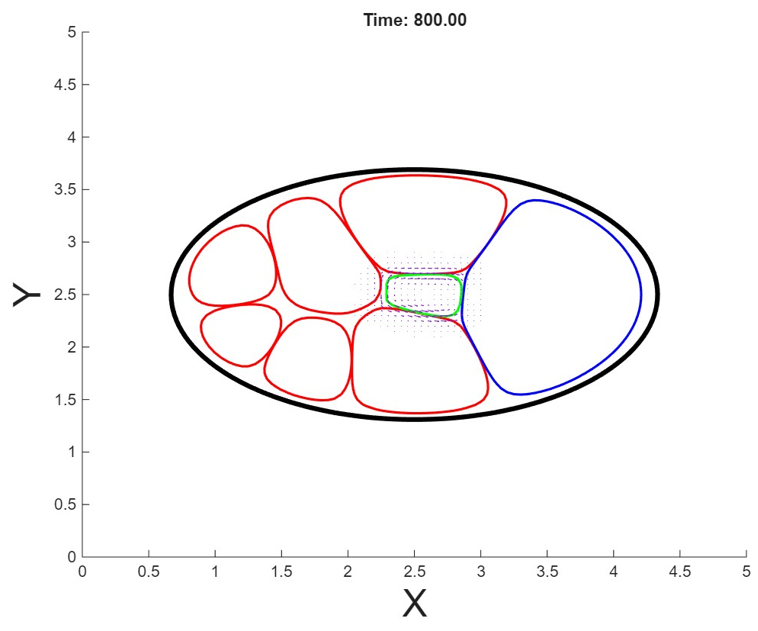}
        }%
    \end{center}
    \caption{%
    {\bf{Border cell cluster migration driven by TIM force under uniform chemoattractant concentration $c(x)$.}} Snapshots at three time points show migration driven by the tangential interface migration (TIM) force (Eq.~\eqref{eq: tim force} in main text). The purple vector field indicates tangential traction generated through contact between the border cell cluster (green) and adjacent nurse cells (red), enabling forward movement even in the absence of spatial heterogeneity in the chemoattractant field.
     }%
   \label{fig:TIM cosh mig}
\end{figure}

In Figure~\ref{fig:TIM cosh mig}(a)-(c), we present snapshots of border cell cluster migration driven solely by the TIM force, as defined in Eq.~\eqref{eq: tim force} of the main text. The purple vector field illustrates the localized tangential interactions between the border cell cluster and adjacent nurse cells. These interactions generate contact-mediated traction that guides the cluster forward, even in the absence of spatial variation in chemoattractant concentration.
To evaluate how extracellular space–dependent chemoattractant profiles influence TIM-driven migration, we compare the cluster’s position and speed under uniform concentration $c(x)$ and spatially varying concentration $c_r(x)$, which accounts for changes in extracellular geometry. 
\begin{figure}[t!]

     \begin{center}
        \subfigure[]{%
            \label{fig:first}
            \includegraphics[width=0.45\textwidth]{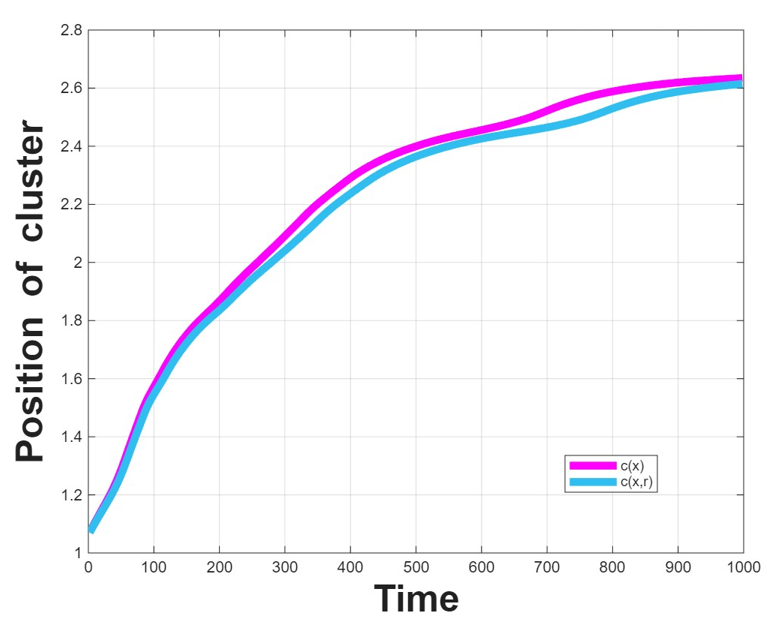}
        }%
        \subfigure[]{%
           \label{fig:second}
           \includegraphics[width=0.435\textwidth]{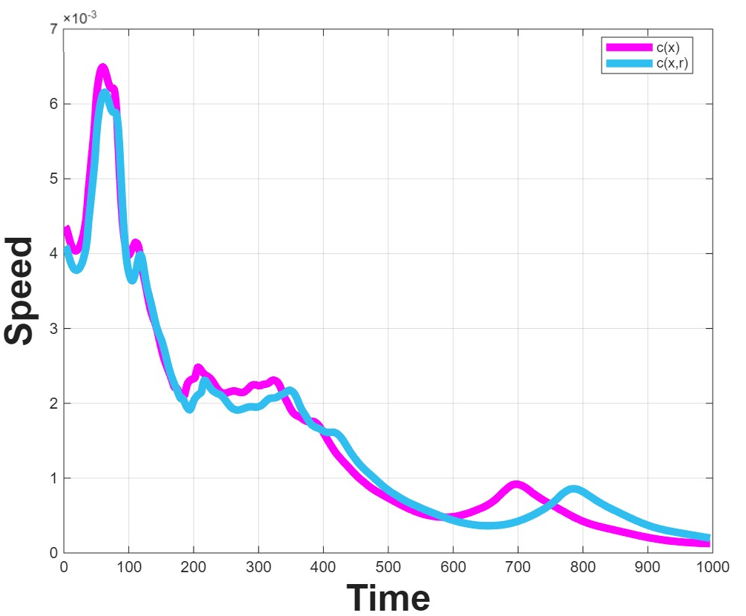}
        }
    \end{center}
    \caption{%
    {\bf{Comparison of TIM force–driven migration under uniform $c(x)$ and spatially varying $c_r(x)$ chemoattractant concentration.}} (a) Position of the border cell cluster over time for simulation using $c(x)$ (magenta) and $c_r(x)$ (cyan). (b) Corresponding cluster speed profiles. While overall migration remains robust under both conditions, subtle differences in speed and trajectory reflect the influence of chemoattractant field geometry on interfacial traction patterns driving the TIM force.
    }%
   \label{fig:TIM cosh and c_r}
\end{figure}

\clearpage

{\bf{Supplemental Movie 1:}} Live border cell migration in a normal genotype. Border cells (green) migrate in a wild-type egg chamber towards the oocyte, as seen by live time-lapse imaging \emph{ex vivo}. Genotype: slbo-Lifeact-GFP; mat$\alpha$-Gal4 / UAS-Myr-tdTomato. Only the GFP is shown. 1 hour and 20 minutes into the movie, the border cells nearly reach the oocyte boundary and then turn to move dorsally (upwards) to be close to the oocyte nucleus, which is the source of the chemoattractant signal Grk. Time stamp = hrs:min. Scale Bar = 50 $\mu$m.
See Figure~\ref{fig:dorsal migration} for a representative time point from this simulation.
\bigskip

{\bf{Supplemental Movie 2:}} Model border cell cluster migration in response to the chemoattractant concentration gradient ($F_{\text{chem}}$). Parameters are set as follows: $\mu_c = 0.045$, $\Gamma = 0.01$, $s = 1$, and $\ell = 0.05$. In this simulation, the cross-sectional area between nurse cells satisfies $A(x) > 0$ and is incorporated into the chemoattractant concentration model.
The dynamics shown here are illustrated in Figure~\ref{fig:chem} (c-d) of the main text.
\bigskip

{\bf{Supplemental Movie 3:}} Model border cell cluster migration in response to the Tangential Interface Migration (TIM) force ($F_{\text{TIM}}$). The arrows indicate tangentially oriented force vectors along the cluster boundary. 
A corresponding snapshot is shown in Figure~\ref{fig:TIM mig} in the main text. 

\bigskip

{\bf{Supplemental Movie 4:}} Dorsal turn of the border cell cluster. The arrow shows the tangential interface migration force, proportional to the local chemoattractant concentration. The simulation corresponds to Figure~\ref{fig:dorsal migration} in the main text. 
\bigskip

{\bf{Supplemental Movie 5:}} Model border cell cluster migration in response to the chemoattractant gradient when the cross-sectional area is fixed at $A(x) = 1$, corresponding to a constant radius $r$. The cluster migrates in response to the chemical gradient.
This simulation provides the context for the results presented in  Figure~\ref{fig:chem cosh mig} in supplementary. 

\bigskip

{\bf{Supplemental Movie 6:}} Migration of the model border cell cluster under the influence of the TIM force when the cross-sectional area is fixed ($A(x) = 1$).
This simulation corresponds to Figure~\ref{fig:TIM cosh mig} in the supplementary.

\end{document}